\documentclass[a4paper,fleqn,usenatbib]{mnras}

\usepackage[T1]{fontenc}
\usepackage{ae,aecompl}
\usepackage{lineno}

\usepackage{graphicx}	
\usepackage{amsmath}	
\usepackage{amssymb}	

\usepackage{hyperref}
\usepackage{cleveref}
\usepackage{subfigure}
\usepackage{float}
\usepackage{amsmath}
\usepackage{etoolbox}
\usepackage{color,soul}
\pretocmd{\abstractname}{\newpage}{}{}

\title[UV-Luminous Star-Forming Hosts of Reddened Quasars]{UV-Luminous, Star-Forming Hosts of z$\sim$2 Reddened Quasars in the Dark Energy Survey}

\author[C. F. Wethers, M. Banerji, P. C. Hewett et al.]{ \parbox{\textwidth}
{C. F. Wethers$^{1}$\thanks{E-mail: cfw36@ast.cam.ac.uk},
M. Banerji$^{1,2}$,
P. C. Hewett$^{1}$,
C. A. Lemon$^{1,2}$,
R. G. McMahon$^{1,2}$,
S. L. Reed$^{1}$,
Y. Shen$^{3,4}$,
F. B. Abdalla$^{5,6}$,
A. Benoit-L{\'e}vy$^{5,7,8}$,
D. Brooks$^{5}$,
E. Buckley-Geer$^{9}$,
D. Capozzi$^{10}$,
A.  Carnero Rosell$^{11,12}$,
M. CarrascoKind$^{13,14}$,
J. Carretero$^{15}$,
C. E. Cunha$^{16}$,
C. B. D'Andrea$^{17}$,
L. N. da Costa$^{11,12}$,
D. L. DePoy$^{18}$,
S. Desai$^{19}$,
P. Doel$^{5}$,
B. Flaugher$^{9}$,
P. Fosalba$^{20}$,
J. Frieman$^{9,21}$,
J. Garc\'ia-Bellido$^{22}$,
D. W. Gerdes$^{23,24}$,
D. Gruen$^{16,25}$,
R. A. Gruendl$^{13,14}$,
J. Gschwend$^{11,12}$,
G. Gutierrez$^{9}$,
K. Honscheid$^{26,27}$,
D. J. James$^{28,29}$,
T. Jeltema$^{30}$,
K. Kuehn$^{31}$,
S. Kuhlmann$^{32}$,
N. Kuropatkin$^{9}$,
M. Lima$^{11,33}$,
M. A. G. Maia$^{11,12}$,
J. L. Marshall$^{18}$,
P. Martini$^{26,34}$,
F. Menanteau$^{13,14}$,
R. Miquel$^{15,35}$,
R. C. Nichol$^{10}$,
B. Nord$^{9}$,
A. A. Plazas$^{36}$,
A. K. Romer$^{37}$,
E. Sanchez$^{38}$,
V. Scarpine$^{9}$,
R. Schindler$^{25}$,
M. Schubnell$^{24}$,
I. Sevilla-Noarbe$^{38}$,
M. Smith$^{39}$,
R. C. Smith$^{29}$,
M. Soares-Santos$^{9}$,
F. Sobreira$^{11,40}$,
E. Suchyta$^{41}$,
G. Tarle$^{24}$,
A. R. Walker$^{29}$;
\small{Affiliations at end of paper}
} 
\\
}

\date{Accepted XXX. Received YYY; in original form ZZZ}

\pubyear{2018}

\begin{document}
\label{firstpage}
\pagerange{\pageref{firstpage}--\pageref{lastpage}}
\maketitle

\begin{abstract}
We present the first rest-frame UV population study of 17 heavily reddened, high-luminosity (E(B-V)$_{\rm{QSO}}\gtrsim$ 0.5; L$_{\rm{bol}}>$ 10$^{46}$ergs$^{-1}$) broad-line quasars at $1.5 < z < 2.7$. We combine the first year of deep, optical, ground-based observations from the Dark Energy Survey (DES) with the near infrared VISTA Hemisphere Survey (VHS) and UKIDSS Large Area Survey (ULAS) data, from which the reddened quasars were initially identified. We demonstrate that the significant dust reddening towards the quasar in our sample allows host galaxy emission to be detected at the rest-frame UV wavelengths probed by the DES photometry. By exploiting this reddening effect, we disentangle the quasar emission from that of the host galaxy via spectral energy distribution (SED) fitting. We find evidence for a relatively unobscured, star-forming host galaxy in at least ten quasars, with a further three quasars exhibiting emission consistent with either star formation or scattered light. From the rest-frame UV emission, we derive instantaneous, dust-corrected star formation rates (SFRs) in the range 25 < SFR$_{\rm{UV}}$ < 365 M$_{\odot}$yr$^{-1}$, with an average SFR$_{\rm{UV}}$ = 130 $\pm$ 95 M$_{\odot}$yr$^{-1}$. We find a broad correlation between SFR$_{\rm{UV}}$ and the  bolometric quasar luminosity. Overall, our results show evidence for coeval star formation and black hole accretion occurring in luminous, reddened quasars at the peak epoch of galaxy formation.    

\end{abstract}



\section{Introduction}
\label{sec:intro}

Active galactic nuclei (AGN) are thought to govern many fundamental processes within galaxies, from the quenching of their star formation to their morphological evolution. In nearby galaxies, tight correlations have been observed between the central black hole mass (M$_{\rm{BH}}$) and several properties of the galactic bulge e.g. bulge mass, M$_{\rm{bulge}}$ \citep{magorrian98}, bulge luminosity, L$_{\rm{bulge}}$ \citep{faber76}, the stellar velocity dispersion, $\sigma$, and the stellar mass, M$_{*}$ \citep{kormendy13}. Furthermore, black hole (BH) accretion and star formation both appear to peak at $z\sim$2 \citep[e.g.][]{aird15, madau14}, indicating a likely link between star formation and AGN activity in galaxies. Some galaxy formation scenarios postulate that the same gas supply can fuel both star formation and accretion onto the BH in massive galaxies (e.g. \citealt{hopkins08, narayanan10}). \cite{sanders88}, for example, suggest that the most luminous systems evolve from merger-driven starbursts to UV-luminous quasars, appearing heavily obscured during the transition phase, when remnant dust from the decaying starburst is being cleared out of the galaxy. Studying the connection between dust obscuration, BH accretion and star formation in the most luminous quasars is therefore an important test of such galaxy evolution models.

Actively accreting super-massive black holes (SMBHs), or quasars, typically outshine their host galaxies by several orders of magnitude, particularly at rest-frame ultraviolet (UV) wavelengths, where the quasar continuum peaks. For this reason, quasar host galaxy studies in the rest-frame UV and optical require highly spatially-resolved images to separate light from the host galaxy from that of the quasar. This limitation means such studies have largely been confined to low redshifts (z < 1), where good spatial resolution is much easier to achieve \citep[e.g.][]{dunlop03,matsuoka15}. At these low redshifts, host galaxy emission has been separated from that of the AGN via both image \citep[e.g.][]{jahnke04,sanchez14} and spectral decomposition \citep[e.g.][]{berk06,matsuoka15}, with these studies finding a strong correlation between M$_{\rm{BH}}$ and M$_{\rm{bulge}}$. However there appears to be no correlation between the quasar accretion rate and star formation rate (SFR) of the host at these low redshifts \citep[e.g.][]{urrutia12}.

At higher redshifts (z$\sim$2) - the epoch at which BH activity peaks - a handful of studies have used high-resolution spaced-based imaging from the \textit{Hubble Space Telescope (HST)} to spatially isolate the host galaxy emission, although these studies have largely targeted moderate-luminosity quasars (L$_{\rm{bol,QSO}}\sim$10$^{43-44}$ergs$^{-1}$), where emission from the galaxy makes up a significant fraction of the flux in the rest-frame UV/ optical \citep[e.g.][]{jahnke04}. In these moderate-luminosity quasars, \cite{jahnke04} find no dependence of the SFR on the quasar luminosity, deriving SFRs $\sim$6\,M$_{\odot}$yr$^{-1}$ (prior to dust correction) across their entire quasar sample. To analyse the hosts of more luminous quasars, studies have primarily relied on observations at longer wavelengths in the far infra-red (FIR) to millimetre regime, where the host galaxy emission peaks \citep[e.g.][]{priddey03,harris16}. Unlike their low-redshift and lower-luminosity counterparts, the SFRs of high-luminosity quasars at z $\gtrsim$ 2 have been shown to correlate with L$_{\rm{bol,QSO}}$, with several studies finding more luminous quasars to reside in more actively star-forming hosts \citep[e.g.][]{coppin08, hatziminaoglou10, rosario12, delvecchio15, xu15, harris16}. 

Several studies have also attempted to probe these luminous, high-redshift quasar hosts in the rest-frame optical to near infra-red (NIR), making use of emission at rest-frame wavelengths of $\sim$1$\mu$m, where the fraction of galaxy-to-quasar emission is larger than in the rest-frame UV \citep[e.g.][]{ridgway01,kukula01}. Even in this wavelength regime, however, these studies have typically required high-resolution space-based imaging from \textit{HST} in order to accurately subtract the quasar point spread function (PSF) from the galaxy emission - a problem which remains challenging \citep[e.g.][]{mechtley16}. The majority of these studies find an enhanced merger fraction in populations of high-luminosity quasars, although this conclusion remains disputed \citep[e.g.][]{villforth16}. 

An alternative approach to observing quasar hosts at optical/ NIR wavelengths is to exploit dust obscuration towards the quasar line-of-sight, which can heavily redden the quasar continuum and enhance the fractional flux contribution of the galaxy at these wavelengths \citep[e.g.][]{urrutia08,urrutia12,glikman15,fan16}. This approach has also been used to study quasar hosts in the rest-frame UV, but the rare nature of these heavily-reddened quasars has limited high-redshift studies to individual objects \citep[e.g.][]{cai14}. These high-redshift, dust-obscured quasars may provide key insights into the connection between quasars and their hosts, with \citet{sanders88} suggesting that a connection between L$_{\rm{bol,QSO}}$ and SFR may be more prominent among these dusty systems than among un-obscured quasars.

This paper presents the first rest-frame UV population study of luminous (L$_{\rm{bol,QSO}}>$10$^{46}$ergs$^{-1}$) heavily dust-reddened quasars at $1.5 < z < 2.7$, selected from the work of \citet{banerji12,banerji15}. The heavy dust reddening in our sample ($E(B-V)\simeq\ $0.5 - 2.0) means the quasar light is obscured in the rest-frame UV, potentially exposing emission from the host galaxy at these wavelengths. However, at $1.5 < z < 2.7$ even the most UV-bright star-forming galaxies with SFR > 100\,M$_{\odot}$yr$^{-1}$ require deep imaging to be detected, in part due to the effects of surface brightness dimming  \citep[e.g.][]{reddy08,reddy12}. With the current generation of optical imaging surveys, such as Hyper Suprime-Cam, VST-KiDS and the Dark Energy Survey (DES), deep, ground-based imaging of these systems is now possible over wide fields (>1000\,deg$^{2}$) for the first time. In this work, we make use of data from the first year of DES observations \citep{drlica17}.

The paper is structured as follows. Section~\ref{sec:data} outlines the sample of dust-reddened quasars to be considered. Section~\ref{sec:sed_modelling} details the methods used to fit spectral energy distributions (SEDs) to the photometry, with the results of the fitting given in Section~\ref{sec:results}. Section~\ref{sec:discussion} compares our results with independent studies of these quasars and our key findings are summarised in Section~\ref{sec:conclusions}. Throughout this work, we assume a flat $\Lambda$CDM cosmology with $H_{0}$ = 70 km s$^{-1}$ Mpc$^{-1}$, $\Omega_{M}$ = 0.3 and $\Omega_{\Lambda}$ = 0.7. All quoted magnitudes are based on the AB system, which is the native magnitude system for DES. In the case of the UKIDSS and VISTA, the following Vega to AB conversions have been applied:

\noindent J$_{\rm{UKIDSS,AB}}$ = J$_{\rm{UKIDSS,Vega}}$+0.938; 

\noindent H$_{\rm{UKIDSS,AB}}$ = H$_{\rm{UKIDSS,Vega}}$+1.379;

\noindent K$_{\rm{UKIDSS,AB}}$ = K$_{\rm{UKIDSS,Vega}}$+1.900;

\noindent J$_{\rm{VISTA,AB}}$  = J$_{\rm{VISTA,Vega}}$+0.937; 

\noindent H$_{\rm{VISTA,AB}}$  = H$_{\rm{VISTA,Vega}}$+1.384; 

\noindent K$_{S\rm{VISTA,AB}}$ = K$_{S\rm{VISTA,Vega}}$+1.839.

\section{Data}
\label{sec:data}

The reddened quasars at redshifts $1.5 < z < 2.7$ outlined in \cite{banerji12,banerji15} form the basis of our investigation. The sample selection criteria and the new imaging data are described below.

\subsection{NIR Selection of Luminous Reddened Quasars}
\label{sec:nir_selection}
Red quasar candidates were identified using NIR imaging from wide-field surveys such as the UKIDSS Large Area Survey (ULAS) and VISTA Hemisphere Survey (VHS), as detailed in \cite{banerji12,banerji15}. Objects that appeared as point-sources in the $K$-band, with \emph{K}$_{AB} <$ 18.4\footnote{Magnitudes were calculated within a 1\,arcsecond radius aperture (\textit{apermag3}) and include an aperture correction appropriate for point sources.}, formed the flux-limited base sample. Targets were further required to possess extremely red NIR colours, (\emph{J}-\emph{K})$_{AB} >$ 1.5 (corresponding to an \emph{E(B-V)} $\gtrsim$ 0.5 at $z\sim 2)$. A point-source restriction was also applied to ensure that the $K$-band light is dominated by an unresolved object, therefore excluding galaxies and isolating high-redshift, NIR-luminous quasars. Of the 66 targets selected in this manner, 61 were successfully followed up with either SINFONI VLT or Gemini-GNIRS observations\citep{banerji12,banerji13,banerji15}. 38 of these were spectroscopically-confirmed to be Type-1 broad-line (BL) quasars at $1.5 < z < 2.7$ and form the parent spectroscopic sample of heavily reddened quasars for this paper. With extinction-corrected bolometric luminosities of L$_{\rm{bol,QSO}}\sim$10$^{47}$ergs$^{-1}$ and M$_{\rm{BH}}$ $\sim$10$^{9-10}$ M$_\odot$ \citep{banerji15}, these quasars are among the most luminous and massive accreting SMBHs known at this epoch.

Detecting even the most luminous, star-forming quasar host galaxies at z $\sim$ 2 requires deep optical imaging data. The most UV-luminous, star-forming galaxies at $z\sim$2 currently known have typical $r$-band magnitudes fainter than $\sim$22 (AB) (e.g. \citealt{reddy08}). These magnitudes lie below the flux limit of wide-field optical imaging surveys such as the Sloan Digital Sky Survey (SDSS), which, until recently was the deepest optical imaging survey available over the $>$ 1000 deg$^2$ area overlapping the VHS and ULAS survey footprints.

\subsection{Reddened Quasars in the Dark Energy Survey (DES)}
\label{sec:des}

We make use of new deep optical photometry for the reddened quasars from observations conducted as part of the Dark Energy Survey (DES). DES is a wide-field survey, imaging 5000 deg$^{2}$ of the southern celestial hemisphere in the \emph{grizY}-bands \citep{frieman13,flaugher15,abbott16}. The 5-year survey began in 2013 and uses the 570 Megapixel DECam on the 4m Blanco telescope at the Cerro Tololo Inter-American Observatory (CTIO). DES is among the deepest wide-field surveys currently in operation and will eventually reach depths of $i$ $\lesssim$ 24.0 (10$\sigma$; AB) for extended sources. 

Here, we primarily make use of observations conducted during the first year of DES operations (2013 Aug - 2014 Feb) corresponding to the Year 1 Annual 1 (Y1A1) internal data release \citep{diehl14}. Sixteen of the 38 reddened quasars in our spectroscopic sample overlap with the DES Y1A1 footprint, which covers $\sim$1800\,deg$^2$. The 10$\sigma$ limiting magnitudes reached by the Y1A1 data in a 2\,arcsec aperture are summarised in Table~\ref{tab:survey_depths} for each of the DES bands.

\begin{table}
	\centering
	\caption{The effective wavelengths ($\lambda_{\rm{eff}}$) and 10$\sigma$ limiting magnitudes reached by the Y1A1 data in a 2\,arcsec diameter aperture.}
	\label{tab:survey_depths}
	\begin{tabular}{ccc} 
		\hline
		\hline
		Filter & $\lambda_{eff}$ (\AA) & \multicolumn{1}{|p{2.5cm}|}{\centering Magnitude Limit \\ (10$\sigma$; AB)} \\
		\hline
		$g$ & 4824 & 23.4 \\ 
		$r$ & 6432 & 23.2 \\ 
        $i$ & 7806 & 22.5 \\ 
        $z$ & 9179 & 21.8 \\ 
        $Y$ & 9883 & 20.1 \\ 
		\hline
	\end{tabular}
\end{table}

\begin{table*}
	\centering
	\caption{Summary of the reddened quasar sample considered in this work. M$_{\rm{BH}}$ and $z$ are derived from the VLT spectral follow-up observations presented in \citet{banerji12, banerji15}. Values of L$_{\rm{bol,QSO}}$ have been calculated based on the results of this paper.}
	\label{tab:datasummary}
	\begin{tabular}{llrcccc} 
		\hline
		\hline
		Name & RA & DEC & $z$ & log$_{10}$(L$_{\rm{bol,QSO}}$/erg s$^{-1}$ (L$_{\odot}$)) & log$_{10}$(M$_{\rm{BH}}$/M$_{\odot}$)\\
		\hline
        ULASJ0016-0038 & 4.0025 & -0.6498 & 2.194 & 46.54 (12.95) & 9.3  \\
        ULASJ1002+0137 & 150.5470 & 1.6185 & 1.595 & 46.51 (12.92) & 10.1  \\
		VHSJ2024-5623 & 306.1074 & -56.3898 & 2.282 & 46.63 (13.05) & 9.8  \\
		VHSJ2028-5740 & 307.2092 & -57.6681 & 2.121 & 47.66 (14.08) & 10.1  \\
		VHSJ2100-5820 & 315.1403 & -58.3354 & 2.360 & 47.08 (13.50) & 9.1  \\
		VHSJ2115-5913 & 318.8817 & -59.2188 & 2.115 & 47.49 (13.91) & 9.3  \\
        ULASJ2200+0056 & 330.1036 & 0.9346 & 2.541 & 47.34 (13.76) & 9.2  \\
		VHSJ2220-5618 & 335.1398 & -56.3106 & 2.220 & 47.87 (14.28) & 9.9  \\
        ULASJ2224-0015 & 336.0392 & -0.2566 & 2.223 & 46.87 (13.28) & 8.9  \\
		VHSJ2227-5203 & 336.9491 & -52.0582 & 2.656 & 46.87 (13.29) & 10.0  \\		
		VHSJ2235-5750 & 338.9331 & -57.8371 & 2.246 & 47.13 (13.55) & 10.1  \\
		VHSJ2256-4800 & 344.1443 & -48.0088 & 2.250 & 47.43 (13.85) & 10.1  \\
		VHSJ2257-4700 & 344.2589 & -47.0156 & 2.156 & 46.60 (13.02) & 9.5  \\
		VHSJ2306-5447 & 346.5010 & -54.7882 & 2.372 & 46.90 (13.32) & 10.0  \\
        ULASJ2315+0143 & 348.9842 & 1.7307 & 2.560 & 48.06 (14.48) & 8.8  \\
		VHSJ2332-5240 & 353.0387 & -52.6780 & 2.450 & 46.78 (13.19) & 9.5  \\
		VHSJ2355-0011 & 358.9394 & -0.1893 & 2.531 & 47.34 (13.75) & 10.1  \\
        \hline
	\end{tabular}
\end{table*}

\begin{figure*}
	\centering  
    \subfigure[ULASJ0016-0038]{\label{fig:colour1}\includegraphics[trim= 100 20 100 40 ,clip,width=0.195\textwidth]{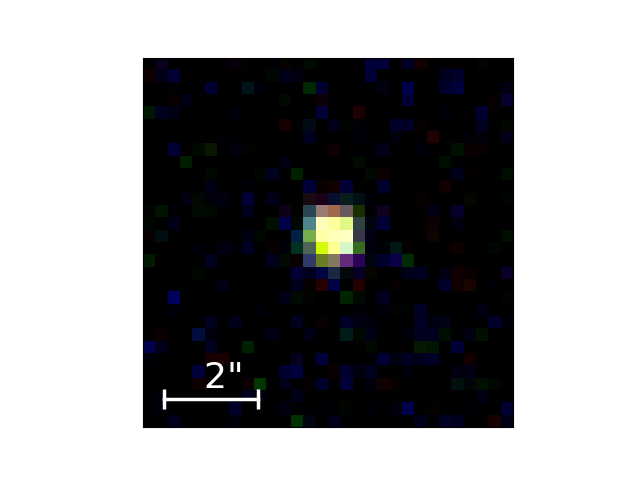}}
    \subfigure[ULASJ1002+0137]{\label{fig:colour2}\includegraphics[trim= 100 20 100 40 ,clip,width=0.195\textwidth]{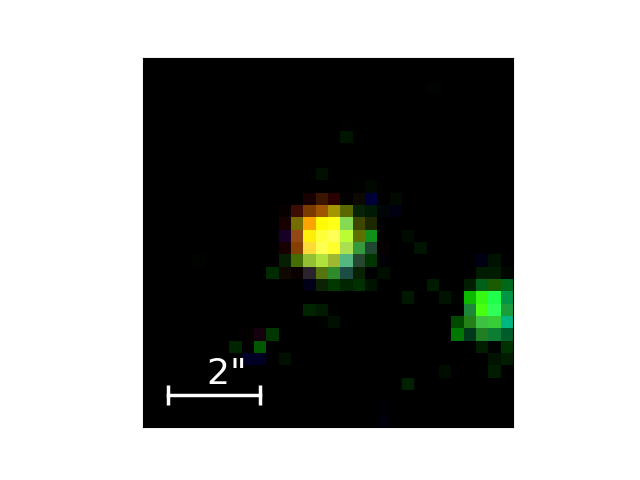}}
	\subfigure[VHSJ2024-5623]{\label{fig:colour3}\includegraphics[trim= 100 20 100 40 ,clip,width=0.195\textwidth]{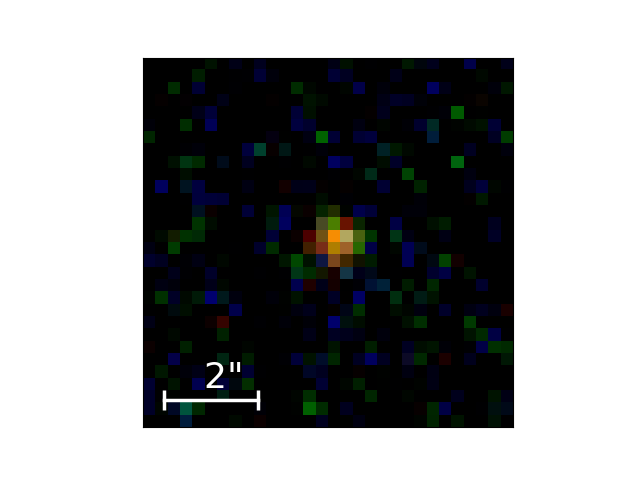}}
    \subfigure[VHSJ2028-5740]{\label{fig:colour4}\includegraphics[trim= 100 20 100 40 ,clip,width=0.195\textwidth]{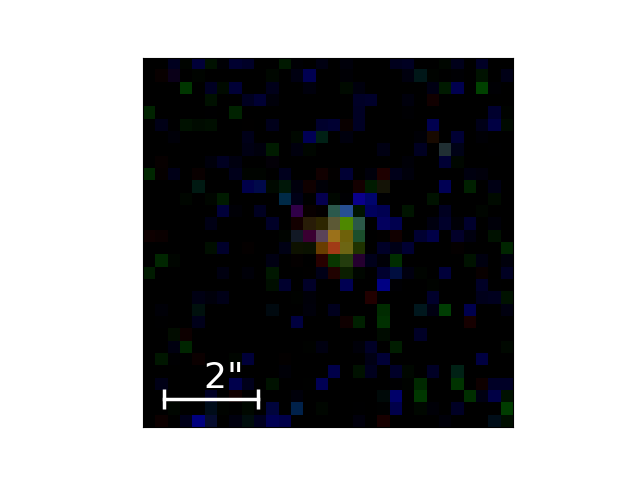}}
	\subfigure[VHSJ2100-5820]{\label{fig:colour5}\includegraphics[trim= 100 20 100 40 ,clip,width=0.195\textwidth]{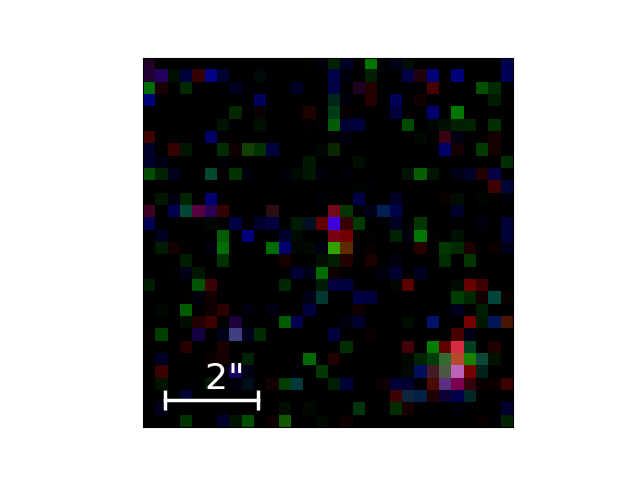}}
    \subfigure[VHSJ2115-5913]{\label{fig:colour6}\includegraphics[trim= 100 20 100 40 ,clip,width=0.195\textwidth]{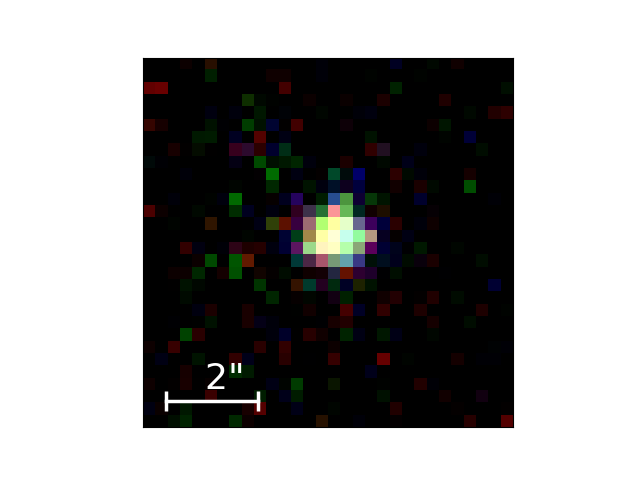}}
	\subfigure[ULASJ2200+0056]{\label{fig:colour7}\includegraphics[trim= 100 20 100 40 ,clip,width=0.195\textwidth]{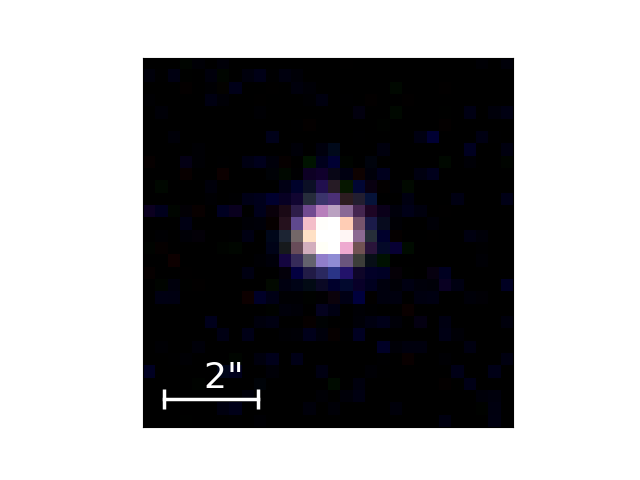}} 
    \subfigure[VHSJ2220-5618]{\label{fig:colour8}\includegraphics[trim= 100 20 100 40 ,clip,width=0.195\textwidth]{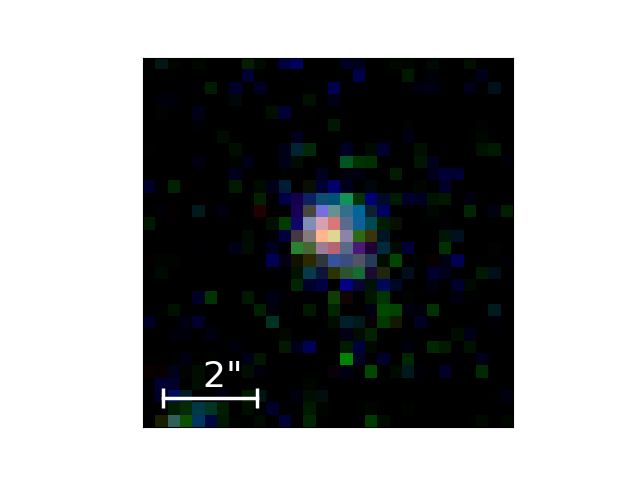}}
    \subfigure[ULASJ2224-0015]{\label{fig:colour9}\includegraphics[trim= 100 20 100 40 ,clip,width=0.195\textwidth]{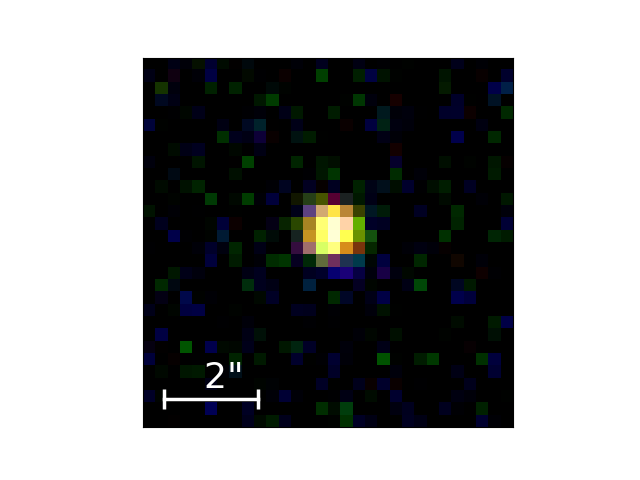}}
    \subfigure[VHSJ2227-5203]{\label{fig:colour10}\includegraphics[trim= 100 20 100 40 ,clip,width=0.195\textwidth]{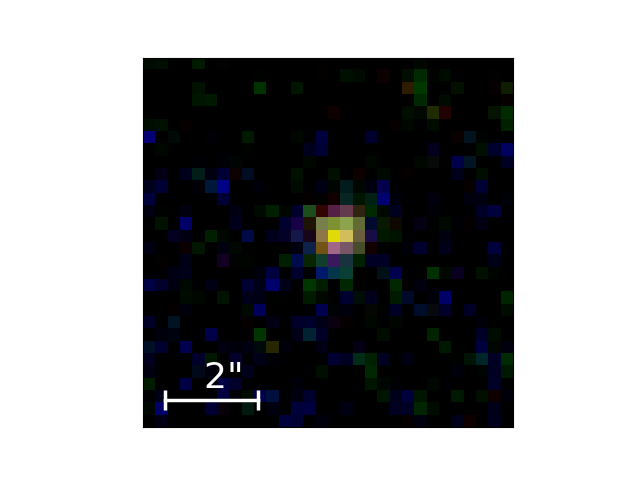}}
    \subfigure[VHSJ2235-5750]{\label{fig:colour11}\includegraphics[trim= 100 20 100 40 ,clip,width=0.195\textwidth]{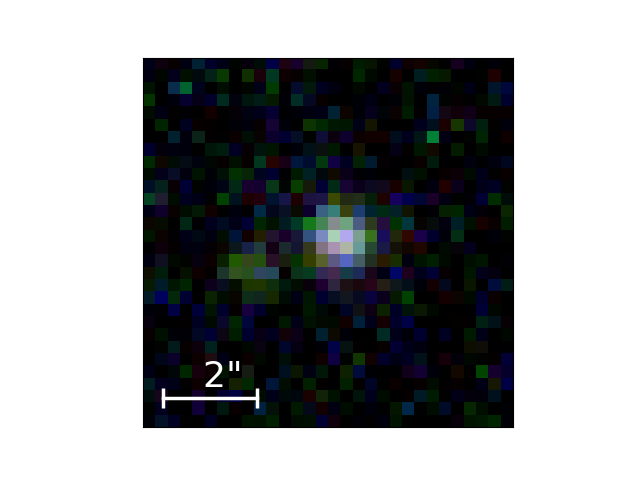}}
    \subfigure[VHSJ2256-4800]{\label{fig:colour12}\includegraphics[trim= 100 20 100 40 ,clip,width=0.195\textwidth]{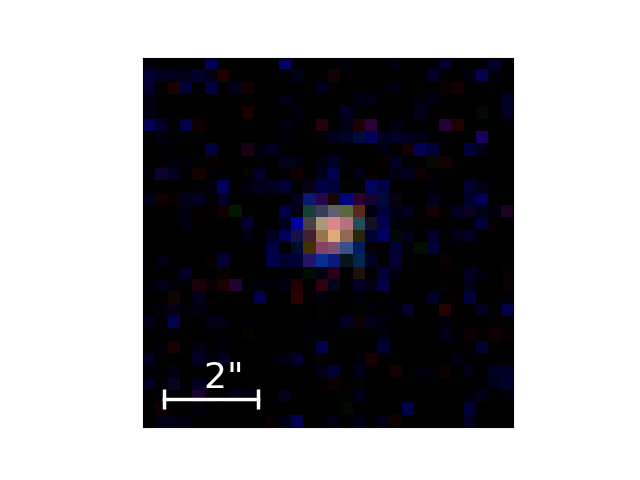}}
	\subfigure[VHSJ2257-4700]{\label{fig:colour13}\includegraphics[trim= 100 20 100 40 ,clip,width=0.195\textwidth]{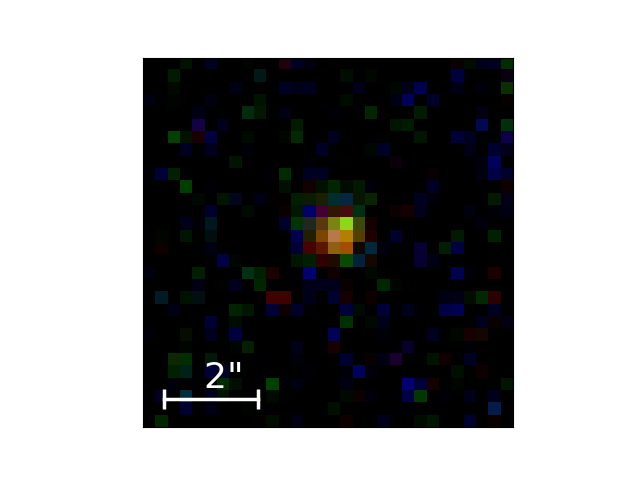}}
	\subfigure[VHSJ2306-5447]{\label{fig:colour14}\includegraphics[trim= 100 20 100 40 ,clip,width=0.195\textwidth]{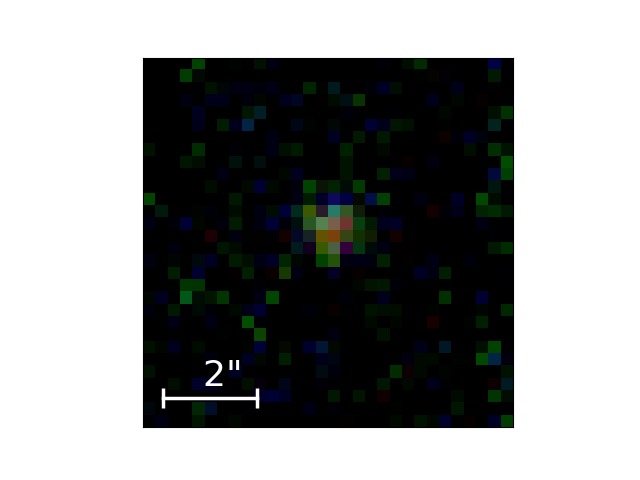}}
	\subfigure[ULASJ2315+0143]{\label{fig:colour15}\includegraphics[trim= 100 20 100 40 ,clip,width=0.195\textwidth]{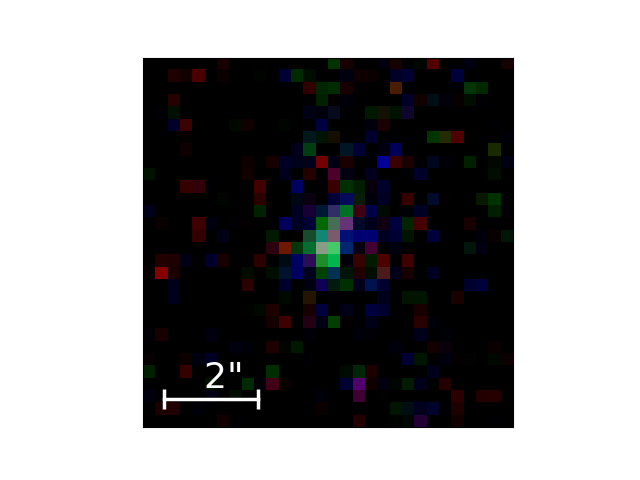}} 
	\subfigure[VHSJ2332-5240]{\label{fig:colour16}\includegraphics[trim= 100 20 100 40 ,clip,width=0.195\textwidth]{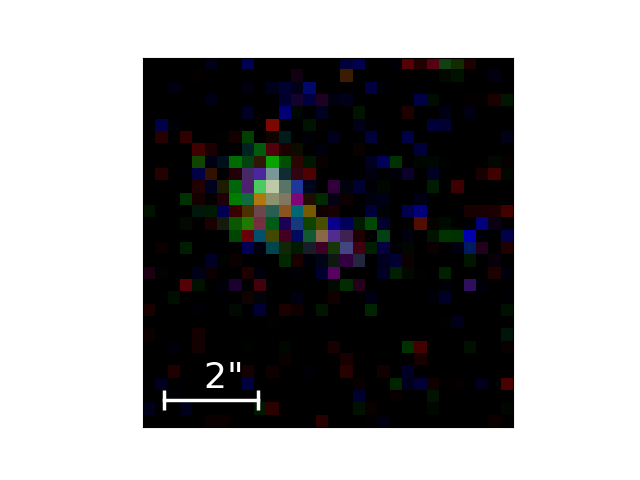}}
	\subfigure[VHSJ2355-0011]{\label{fig:colour17}\includegraphics[trim= 100 20 100 40 ,clip,width=0.195\textwidth]{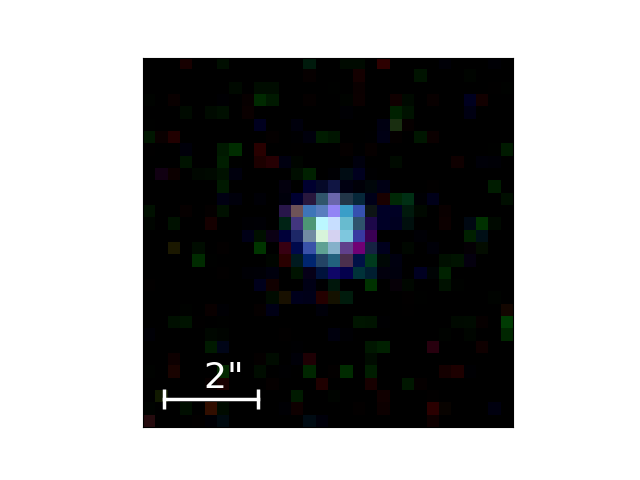}}
\caption{Composite DES colour images, $g$ (blue), $r$ (green), $i$ (red), for the quasar sample.}
\label{fig:colour_images}
\end{figure*}

For a single quasar, ULASJ1002+0137, in the COSMOS field, we also make use of DECam observations conducted as part of the Science Verification. This data goes considerably deeper than the Y1A1 data, reaching an $i$-band magnitude limit of 25.1 in a 2\,arcsec aperture (10$\sigma$; AB). We note however, that the optical magnitudes of ULASJ1002+0137 are such that it would also be detected with the Y1A1 depths, had this region been part of the Y1A1 footprint. The sample considered in this work therefore consists of 17 spectroscopically-confirmed reddened quasars at $1.5<z<2.7$, shown to be representative of the parent sample of 38 reddened quasars in terms of redshift, luminosity and BH mass. The DECam $grizY$ band observations for the sample are summarised in Table~\ref{tab:datasummary}, with the corresponding $gri$-band colour composite images given in Fig.~\ref{fig:colour_images}.

At redshifts $1.5<z<2.7$, the DES filters trace emission at rest-frame UV wavelengths ($\lambda_{\rm{rest}}\sim$1300 - 3800\,\AA). We use the DES \textsc{mag$\_$auto} SExtractor magnitudes for our sources, which have been shown to perform best in SED-fitting of galaxies with the DES data (e.g. \citealt{sanchez14}). The \textsc{mag$\_$auto} magnitudes are preferred to the DES model magnitudes as the model magnitudes assume an exponential disk model, which may not be appropriate for our high-redshift galaxies. We note however, that differences between the model and \textsc{mag$\_$auto} magnitudes are typically small and our choice of magnitudes therefore does not affect the main conclusions of this study.  The \textsc{mag$\_$auto} magnitudes do however underestimate the fluxes of point sources, as no aperture correction is applied. Indeed, \citet{reed17} find that for point sources in DES Y1A1, the \textsc{mag$\_$auto} values are systematically fainter by $\sim$0.1 mag compared to the \textsc{mag$\_$psf} values. Although our quasars are mostly extended in the DES $griz$-bands, they are often unresolved in the DES $Y$-band, where the \textsc{mag$\_$auto} values are $\sim$0.1 mags fainter than the \textsc{mag$\_$psf} values. Once again, we find these magnitude variations to have a negligible effect on the results of this study. The DES \textsc{mag$\_$auto} magnitudes for all 17 sources are presented in Table~\ref{tab:photometry1}. All photometry has been corrected for any spatial non-uniformity in the global calibration and for the effects of Galactic extinction using the stellar locus regression technique described in \cite{drlica17}. Sixteen of the 17 quasars in our sample are detected at $\geq 10\sigma$ in at least one of the $gri$ DES bands. 

\begin{table*}
	\centering
	\caption{DES \textsc{mag$\_$auto} magnitudes for the reddened quasar sample. All magnitudes are on the AB system and have been corrected for the Stellar Locus Regression (SLR) offsets in the  \emph{g,r,i} and \emph{z}-bands. $g_{\rm{QSO,SED}}$ denotes the predicted $g$-band magnitude, based on the SED fits to the NIR photometry alone \citep{banerji12,banerji15}.}
	\label{tab:photometry1}
	\begin{tabular}{lccccccccc} 
		\hline
		\hline
		Name & $g_{\rm{QSO,SED}}$ & $g_{\rm{auto}}$ & $g_{\rm{QSO,SED}}$ - $g_{\rm{auto}}$ & $r_{\rm{auto}}$ & $i_{\rm{auto}}$ & $z_{\rm{auto}}$ & $Y_{\rm{auto}}$ \\
		\hline
        ULASJ0016-0038 & 24.36 & 23.16 $\pm$0.08 & 1.20 & 21.80 $\pm$0.04 & 21.19 $\pm$0.04 & - 			  & 20.43 $\pm$0.10 \\
        ULASJ1002+0137 & 28.14 & 23.55 $\pm$0.05 & 4.59 & 23.01 $\pm$0.03 & 22.14 $\pm$0.02 & 21.36 $\pm$0.02 & 21.06 $\pm$0.04 \\
		VHSJ2024-5623  & 25.34 & 24.81 $\pm$0.37 & 0.53 & 23.68 $\pm$0.16 & 22.34 $\pm$0.07 & 21.63 $\pm$0.07 & 21.69 $\pm$0.23 \\
		VHSJ2028-5740  & 29.19 & 24.15 $\pm$0.24 & 5.04 & 23.32 $\pm$0.12 & 22.45 $\pm$0.08 & 21.61 $\pm$0.08 & 21.37 $\pm$0.26 \\
		VHSJ2100-5820  & 26.95 & 25.37 $\pm$0.49 & 1.58 & 24.51 $\pm$0.39 & 23.56 $\pm$0.21 & 22.17 $\pm$0.13 & 22.37 $\pm$0.68 \\
		VHSJ2115-5913  & 27.61 & 22.52 $\pm$0.05 & 5.09 & 22.28 $\pm$0.08 & 21.90 $\pm$0.09 & 21.14 $\pm$0.11 & 21.37 $\pm$0.50 \\
        ULASJ2200+0056 & 23.70 & 22.03 $\pm$0.03 & 1.67 & 21.36 $\pm$0.02 & 20.71 $\pm$0.02 & 19.90 $\pm$0.01 & 19.60 $\pm$0.04 \\
		VHSJ2220-5618  & 24.99 & 23.24 $\pm$0.06 & 1.75 & 22.63 $\pm$0.06 & 21.72 $\pm$0.04 & 20.53 $\pm$0.03 & 20.04 $\pm$0.05 \\
        ULASJ2224-0015 & 24.78 & 23.65 $\pm$0.10 & 1.13 & 22.82 $\pm$0.06 & 21.77 $\pm$0.04 & 20.89 $\pm$0.03 & 21.03 $\pm$0.11 \\
		VHSJ2227-5203  & 28.59 & 23.88 $\pm$0.16 & 4.71 & 23.22 $\pm$0.08 & 22.45 $\pm$0.05 & 21.35 $\pm$0.05 & 21.56 $\pm$0.18 \\	
		VHSJ2235-5750  & 24.44 & 23.20 $\pm$0.09 & 1.24 & 22.71 $\pm$0.07 & 21.82 $\pm$0.05 & 21.05 $\pm$0.04 & 20.74 $\pm$0.09 \\
		VHSJ2256-4800  & 24.27 & 22.76 $\pm$0.07 & 1.51 & 22.55 $\pm$0.05 & 22.20 $\pm$0.08 & 21.32 $\pm$0.05 & 20.90 $\pm$0.15 \\
		VHSJ2257-4700  & 25.88 & 25.57 $\pm$0.46 & 0.31 & 23.83 $\pm$0.14 & 22.48 $\pm$0.07 & 21.57 $\pm$0.07 & 21.43 $\pm$0.26 \\
		VHSJ2306-5447  & 25.87 & 24.07 $\pm$0.12 & 1.80 & 23.90 $\pm$0.16 & 22.66 $\pm$0.08 & 21.69 $\pm$0.06 & 21.75 $\pm$0.34 \\
        ULASJ2315+0143 & 30.55 & 22.70 $\pm$0.08 & 7.85 & 22.46 $\pm$0.10 & 22.47 $\pm$0.20 & 21.94 $\pm$0.19 & - 				\\
		VHSJ2332-5240  & 25.78 & 23.49 $\pm$0.10 & 2.29 & 23.49 $\pm$0.14 & 23.01 $\pm$0.15 & 22.27 $\pm$0.14 & 22.54 $\pm$0.82 \\
		VHSJ2355-0011  & 25.98 & 22.43 $\pm$0.03 & 3.55 & 22.51 $\pm$0.06 & 22.18 $\pm$0.06 & 21.64 $\pm$0.06 & 21.41 $\pm$0.19 \\
        \hline
	\end{tabular}
\end{table*}

\begin{table*}
	\centering
	\caption{\textsc{modest} star-galaxy classifiers for the quasar sample, where 0 = unphysical PSF fit (likely star), 1 = high confidence galaxy, 2 = high confidence star and 3 = ambiguous.}
	\label{tab:modest}
	\begin{tabular}{lcccc} 
		\hline
		\hline
		Name & \textsc{modest}$_{\rm{g}}$ & \textsc{modest}$_{\rm{r}}$ & \textsc{modest}$_{\rm{i}}$ & \textsc{modest}$_{\rm{z}}$ \\
		\hline
        ULASJ0016-0038 & 2 & 2 & 2 & 1 \\
        ULASJ1002+0137 & 1 & 1 & 1 & 3 \\
		VHSJ2024-5623  & 1 & 1 & 3 & 3 \\
		VHSJ2028-5740  & 1 & 1 & 1 & 3 \\
		VHSJ2100-5820  & 1 & 1 & 1 & 1 \\
		VHSJ2115-5913  & 1 & 1 & 1 & 3 \\
        ULASJ2200+0056 & 2 & 2 & 2 & 2 \\
		VHSJ2220-5618  & 1 & 1 & 1 & 1 \\
        ULASJ2224-0015 & 0 & 2 & 2 & 2 \\
		VHSJ2227-5203  & 3 & 2 & 3 & 0 \\		
		VHSJ2235-5750  & 2 & 1 & 1 & 3 \\
		VHSJ2256-4800  & 1 & 1 & 3 & 3 \\
		VHSJ2257-4700  & 0 & 3 & 3 & 2 \\
		VHSJ2306-5447  & 1 & 1 & 1 & 3 \\
        ULASJ2315+0143 & 1 & 1 & 1 & 1 \\
		VHSJ2332-5240  & 1 & 1 & 0 & 2 \\
		VHSJ2355-0011  & 3 & 1 & 1 & 2 \\
        \hline
	\end{tabular}
\end{table*}

The DES \textsc{mag$\_$auto} values presented in Table~\ref{tab:photometry1} appear to over-estimate the $g$-band flux compared to those found via the SED fitting in \cite{banerji12,banerji15} ($g_{\rm{QSO,SED}}$). Assuming the reddening towards the quasar sight line calculated from modelling the NIR photometry \citep{banerji12,banerji15}, we derive $g$-band magnitudes $>$1.0 mag fainter than the DES \textsc{mag$\_$auto} magnitudes in 15 of the 17 quasars. This suggests that the DECam images are detecting some source of excess emission in the rest-frame UV that cannot be accounted for by a reddened quasar template.

To quantify whether the excess emission in the $g$-band photometry (Table~\ref{tab:photometry1}) extends beyond the local PSF, we use the \textsc{modest} star-galaxy classifier \citep{drlica17}, based on the \textsc{spread\_model} quantity from SExtractor. \textsc{spread\_model} is defined to be a normalised linear discriminant between the best-fit local PSF-model and a more extended, exponential disk model convolved with the PSF \citep{desai12}. The \textsc{modest} classifier takes account both of the value of \textsc{spread\_model} and its associated error in each band. \textsc{modest} has been optimised for the selection of a high-purity galaxy sample in DES by comparing to classifications from both deep space-based (\textit{HST}) and high-quality ground-based (CFHTLens) surveys. The contamination rate from stars is estimated to be $<$ 5 per cent for $i < 22.5$ \citep{drlica17}. \textsc{modest} classifications for the full sample are given in Table~\ref{tab:modest}, where a \textsc{modest} classification of 0 denotes an unphysical PSF fit (likely star) while 1 indicates a high-confidence galaxy, 2 indicates a high-confidence star and 3 denotes an ambiguous classification. A detailed morphological analysis is beyond the scope of this paper. Instead, we simply seek to discriminate between point-like sources and extended images, which could indicate the detection of the host galaxy.

Ten of the 17 reddened quasars in our sample are morphologically classified as high-confidence galaxies in the DES $g$-band, with a further two having an ambiguous classification and therefore also consistent with being spatially extended. This number of high-confidence galaxies increases to 12 in the $r$-band, representing 70\% of our sample. Even in the redder DES bands, 9 (7) of the quasars are classified as being extended beyond the local PSF in the $i$- ($z$-)bands. Quasar emission is spatially unresolved, yet the \textsc{modest} classifications in Table~\ref{tab:modest} indicate a significant fraction of our quasar sample to be spatially resolved in the DES ground-based imaging. Moreover, this fraction of extended sources increases at bluer wavelengths. Investigating the reason for the spatial extension and the excess rest-frame UV emission seen from these quasars in the DES data forms the central aim of the paper.

\section{Spectral Energy Distribution (SED) Fitting}
\label{sec:sed_modelling}

To characterise the nature of the rest-frame UV emission seen in the DECam images, we fit SED models to the DES $grizY$ and the ULAS/VHS NIR $JHK$ photometry for each source. We thus determine a model SED spanning rest-frame UV through optical wavelengths (approximately 1400-7000\,\AA) for each of the reddened quasars in our sample. In cases where the photometric errors are $<$ 10 per cent of the observed flux, an error floor of 10 per cent is applied prior to the SED-fitting to mitigate against the effect of any systematic uncertainties in the photometry and quasar variability (Section~\ref{sec:qso_variability}). This floor has been selected to represent the scatter in unreddened quasar SEDs over the rest-frame wavelength interval 1200$-$10\,000\,\AA\ . Although the parametric quasar model used in this work (Section~\ref{sec:qso_template}) is found empirically to reproduce the broad-band colours ($ugrizYJHK$) of unreddened quasars \footnote{The quasar model reproduces broad-band quasar colours to $\sigma\simeq 0.1$\,mag, based on quasars at $0.2 < z < 4.0$ with $m_i<19.1$ from the SDSS DR7 quasar catalogue} \citep{schneider10}, not all quasar SEDs look exactly the same, rather there appears to be a scatter $\sim$ 10 per cent. The effectiveness of the parametric model at reproducing the broad-band quasar colours is not expected to improve among reddened populations and so we apply this scatter as a floor on the photometric errors to avoid large contributions to the $\chi^2$ values (Section~\ref{sec:fitting_uv}).

\subsection{Quasar Model}
\label{sec:qso_template} 

We make use of a refined version of the quasar model described in \citet{maddox08}, which has been used in a number of studies including \cite{hewett06} and \citet{maddox12}. The continuum in the rest-frame UV is represented by two power-laws, $f(\nu)$ $\propto$ $\nu^{\alpha}$, where $\alpha = 0.424$ at $\lambda$ $\leq$ 2340\,\AA\ and $\alpha = -0.167$ at $\lambda$ > 2340\,\AA. To account for the reddening of the quasar, we apply an extinction curve, which has been determined empirically using quasars in the seventh data release of the Sloan Digital Sky Survey (SDSS DR7) and is very similar in form to that presented by \cite{gallerani10}. Whilst there is no 2200\,\AA \ feature in the extinction curve, the amount of extinction at wavelengths < 2500\,\AA \ is somewhat less than that derived from the Small Magellanic Cloud (SMC) \citep{pei92}. We note, however, that due to the significant reddening of the quasars in our sample ($E(B-V)>0.5$), the extinction of the quasar light in the rest-frame UV is so great that the exact form of the extinction curve is not important. The results presented in this paper would be essentially identical if an SMC-like extinction curve were used.

The above quasar model also incorporates UV and optical emission lines and FeII multiplets based on the Large Bright Quasar Survey (LBQS) composite of \cite{francis91}. In general the strengths of the emission lines do not significantly affect the quasar broad-band colours except in the case of H-$\alpha$, which has a higher equivalent width relative to the other lines. We therefore adjust the H-$\alpha$ equivalent width in the quasar model for each reddened quasar to match that measured from the NIR spectra in \cite{banerji12,banerji15}. Hence, all reddening estimates for the quasars presented in this paper account for the effect of the H$\alpha$ emission line on the observed colours.

\subsection{Host Galaxy Model}
\label{sec:gal_template} 

A key aim of this study is to investigate whether the emission observed in the rest-frame UV is consistent with that of a star-forming host. Given the limited photometry tracing the rest-frame UV emission, we are unable to solve for all properties of the galaxy in the fitting. Instead, we consider a single star-forming galaxy SED template from \citet{bruzual03} with a constant star formation history (SFH) to model the host galaxy emission. The galaxy model is attenuated using a two-component \citet{charlot00} dust model, with $\tau_{\rm{V}}$ = 1.0 (corresponding to an $E(B-V)_{\rm{gal}} \simeq$ 0.35) affecting the young stellar populations, which is consistent with the level of dust extinction seen in local star-forming galaxies. In cases where we see a significant deviation of the rest-frame UV-slope as traced by our $g$ and $r$-band photometry, from that in the default host galaxy model, we also consider galaxy models with $\tau_{\rm{V}}$ = 0.2 and $\tau_{\rm{V}}$ = 5.0, representing bluer and redder rest-frame UV colours respectively \citep{bruzual03}. We note however, that the rest-frame UV slope in the host galaxy model is sensitive to both the dust content ($\tau_{\rm{V}}$) and the SFH. Due to the limited photometry tracing the rest-frame UV and the degenerate nature of these two parameters, we cannot derive meaningful estimates for both $\tau_{\rm{V}}$ and the SFH, so we are limited to these discrete scenarios. In all cases, we assume a \citet{chabrier03} initial mass function (IMF) and solar metallicity.

\subsection{Fitting Method}
\label{sec:fitting_uv}

We fit the combined optical+NIR photometry of the individual quasars using combinations of the models described above, in order to investigate the source of the excess rest-frame UV emission. In particular, we consider three scenarios: (i) the rest-frame UV flux arises from resonantly scattered Lyman-$\alpha$ emission, (ii) the excess emission is scattered continuum light from the central quasar, and (iii) we are seeing UV emission due to star formation in the quasar host galaxy. Although in reality, the rest-frame UV emission is likely due to some combination of the above, the limited DES photometry means we are unable to simultaneously solve for a mix of multiple scenarios. Instead, we seek only to determine the dominant source in each case. To quantify the likelihood of each scenario, we make use of $\chi^{2}$ statistics, calculating the reduced $\chi^{2}$ for each quasar in the sample, i.e.

\begin{equation}
	\chi^{2}_{\rm{red}} = \sum_{i} \bigg( \frac{F_{\rm{phot,i}} - F_{\rm{mod,i}}}{\sigma_{\rm{i}}}\bigg)^{2} \bigg/ N, 
	\label{eq:atten1}
\end{equation}

\noindent where a $\chi^{2}_{\rm{red}} \sim$ 1 indicates a good fit. $F_{\rm{phot}}$ and $F_{\rm{mod}}$ denote the photometric and model fluxes respectively. $N$ is the number of degrees of freedom (with $N$ = $N_{\rm{data}}-N_{\rm{param}}$) and $\sigma$ denotes the photometric errors, floored at 10 per cent. In order to obtain full posterior distributions for all free parameters and marginalise over nuisance parameters, we employ a Markov Chain Monte-Carlo (MCMC) method \citep{metropolis53,hastings70,foreman13} to explore the parameter space. Throughout the fitting we assume flat priors across the entire parameter space.

\section{Results}
\label{sec:results}

This section presents the results of fitting different combinations of the above models to the combined optical+NIR photometry. We initially consider only a reddened quasar template, but later consider three possible sources of the rest-frame UV emission - (i) resonantly-scattered Lyman-$\alpha$, (ii) scattered quasar continuum and (iii) a relatively un-obscured star-forming host galaxy.

\subsection{Quasar Only SED-Fits}
\label{sec:quasar_only}

\begin{figure*}
\begin{tabular}{cc}
	\centering
    \subfigure[ULASJ0016-0038]{\label{fig:0038_agn}\includegraphics[trim=20 -5 200 40,clip,width=0.35\textwidth]{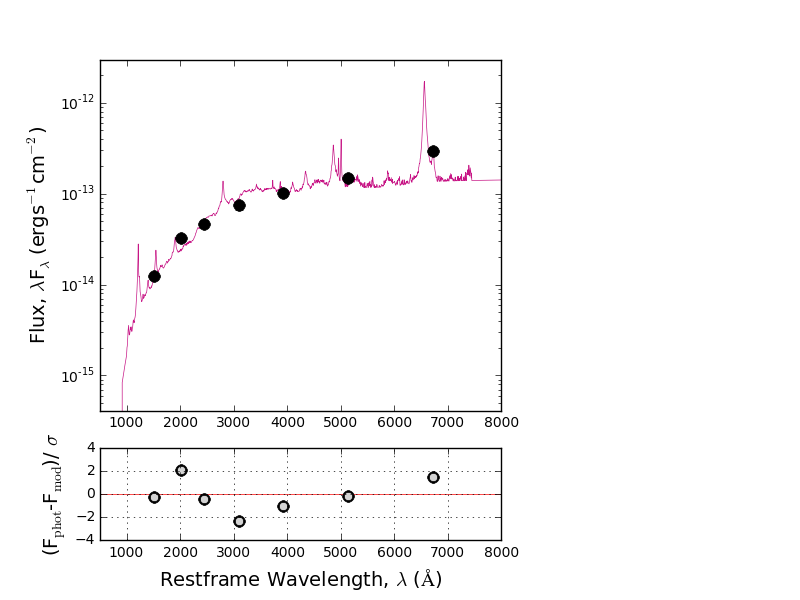}} &
	\subfigure[VHSJ2024-5623]{\label{fig:5623_agn}\includegraphics[trim=20 -5 200 40,clip,width=0.35\textwidth]{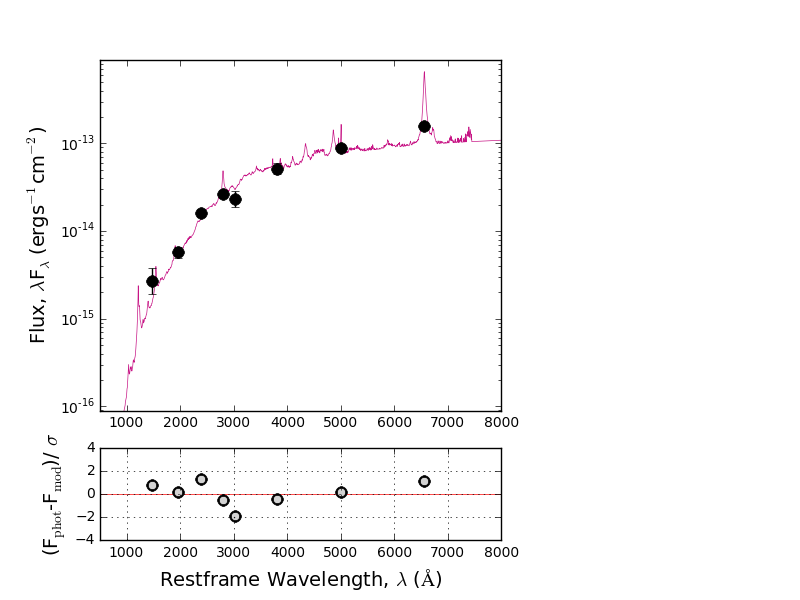}} \\ 
    \subfigure[VHSJ2100-5820]{\label{fig:5820_agn}\includegraphics[trim=20 -5 200 40,clip,width=0.35\textwidth]{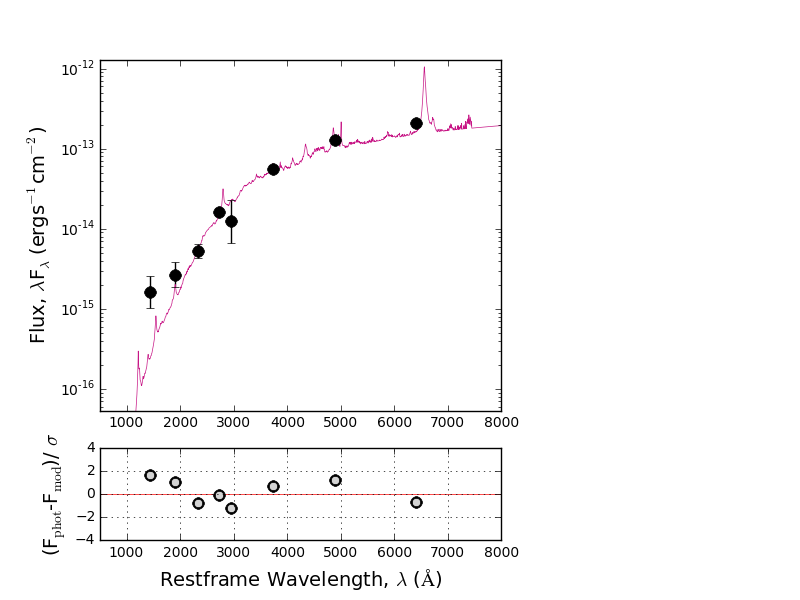}} &
	\subfigure[VHSJ2257-4700]{\label{fig:4700_agn}\includegraphics[trim=20 -5 200 40,clip,width=0.35\textwidth]{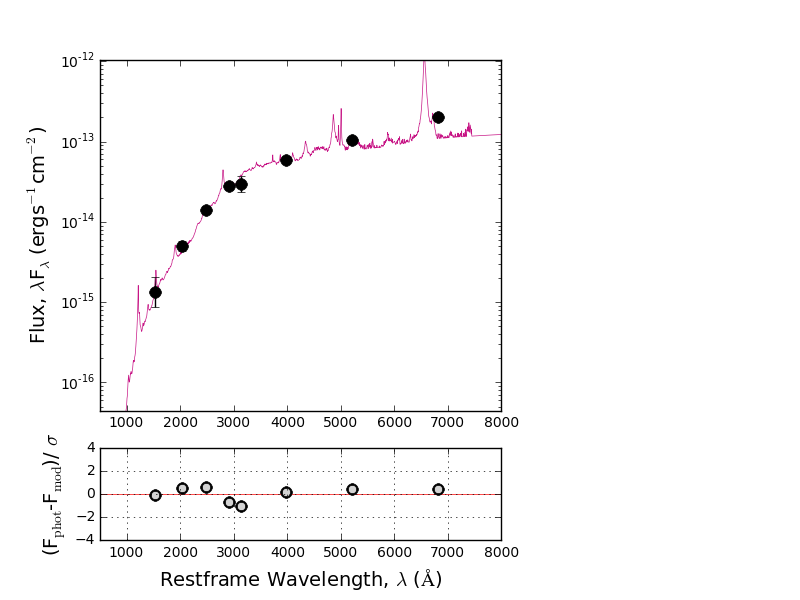}} 
\end{tabular}
\caption{\emph{Upper panels:} Best-fit quasar SEDs for ULASJ0016-0038 (a), VHSJ2024-5623 (b), VHSJ2100-5820 (c), VHSJ2257-4700 (d) for which $\chi^{2}_{\rm{red,QSO}}$ < 2.5. Black points show the $grizY$ DES and $JHK$ VHS/ULAS photometry. \emph{Lower panels:} The error-weighted residuals for each fit.}
\label{fig:qso_only}
\end{figure*}

We begin by modelling the photometry with a reddened quasar template, setting the only free parameter - the quasar dust reddening ($E(B-V)_{\rm{QSO}}$) - to lie in the range 0.3 $\leq E(B-V)_{\rm{QSO}} \leq$ 5.0. Four quasars are found to be well-fit by this model - VHSJ2024-5623, VHSJ2100-5820, VHSJ2257-4700 and ULASJ0016-0038, returning a $\chi^{2}_{\rm{red,QSO}}$ < 2.5 and residuals ($F_{\rm{mod}}-F_{\rm{phot}}$) $<2\sigma$ across all bands. The fitted templates for each of these four quasars, along with their associated residuals, are presented in Fig.~\ref{fig:qso_only}.

The derived values of $E(B-V)_{\rm{QSO}}$ for each of the four quasars in Fig.~\ref{fig:qso_only} are presented in Table~\ref{tab:ebv}, with all four quasars returning an $E(B-V)_{\rm{QSO}}$ consistent to within 0.1 of those presented in \cite{banerji12,banerji15} (hereafter denoted $E(B-V)_{\rm{QSO,B}}$). Although values of $E(B-V)_{\rm{QSO,B}}$ were derived using only the NIR $JHK$-band photometry, our fitting provides little evidence for additional rest-frame UV emission in the DES Y1A1 photometry of these quasars. Despite the \textsc{modest} classifier characterising two of these sources (VHSJ2024-5623 and VHSJ2100-5820) as high-confidence galaxies in the DES $g$ and $r$-bands, we find that a reddened quasar SED is sufficient to fit the photometric data in both cases. Any extended emission in these galaxies is therefore faint and difficult to disentangle from the quasar emission via SED modelling. As such, we find these four quasars to be well characterised by a reddened-quasar template alone and therefore exclude them from the remainder of our analysis. Conversely, the remaining 13 quasars in the sample return $\chi^{2}_{\rm{red,QSO}}$ > 5 when fitted in the above manner, indicating that some additional component is required in the model to account for the observed rest-frame UV flux.

\begin{table}
	\centering
	\caption{$E(B-V)_{\rm{QSO}}$ values for the four quasars returning $\chi^{2}_{\rm{red,QSO}}$ < 2.5. Dust extinction values derived by fitting the NIR photometry \citep{banerji12, banerji15} ($E(B-V)_{\rm{QSO,B}}$) are given for reference.}
	\label{tab:ebv}
	\begin{tabular}{lcc}
		\hline
		\hline
		Name & $E(B-V)_{\rm{QSO}}$ & $E(B-V)_{\rm{QSO,B}}$\\
        \hline  
        ULASJ0016-0038 & 0.43 & 0.5 \\[3pt]
		VHSJ2024-5623 & 0.58 & 0.6 \\[3pt]
		VHSJ2100-5820 & 0.81 & 0.8 \\[3pt]
		VHSJ2257-4700 & 0.67 & 0.7 \\[3pt]
		\hline
	\end{tabular}
\end{table}

\subsection{Scattered Light}
\label{sec:uv_emission}

\subsubsection{Scattered Lyman-alpha Emission}
\label{sec:lyman_alpha} 

One possible explanation for the rest-frame UV flux in the 13 remaining quasars is Lyman-$\alpha$ emission, which can appear spatially extended due to resonant scattering by the surrounding material. We immediately rule out this possibility for seven of the quasars in our sample, as the Lyman-$\alpha$ emission falls outside the wavelength range of the DES $g$-band\footnote{Lyman-$\alpha$ lies at 1216\,\AA\ in the rest-frame, which corresponds to 3155-4446\AA\ in the observed frame over the full redshift range of our sample. Thus the Lyman-$\alpha$ line only affects the DES $g$-band fluxes in our SED fits for all the quasars.}. We model the remaining six quasars - ULASJ2200+0056, VHSJ2227-5203, VHSJ2306-5447, ULASJ2315+0143, VHSJ2332-5240 and VHSJ2355-0011 - with a two-component model consisting of a reddened quasar template and a Lyman-$\alpha$ emission line. As in the quasar-only fitting (Section~\ref{sec:quasar_only}), $E(B-V)_{\rm{QSO}}$ is set to lie in the range 0.3 $\leq E(B-V)_{\rm{QSO}} \leq$ 5.0, while the fraction of scattered Lyman-$\alpha$ can take values 0 $\leq f_{\rm{ly\alpha}}\leq$ 1, where $f_{\rm{ly\alpha}}$ = 1 represents the complete scattering of Lyman-$\alpha$ photons.

\begin{figure}
	\includegraphics[trim=20 0 150 40,clip,width=0.5\textwidth]{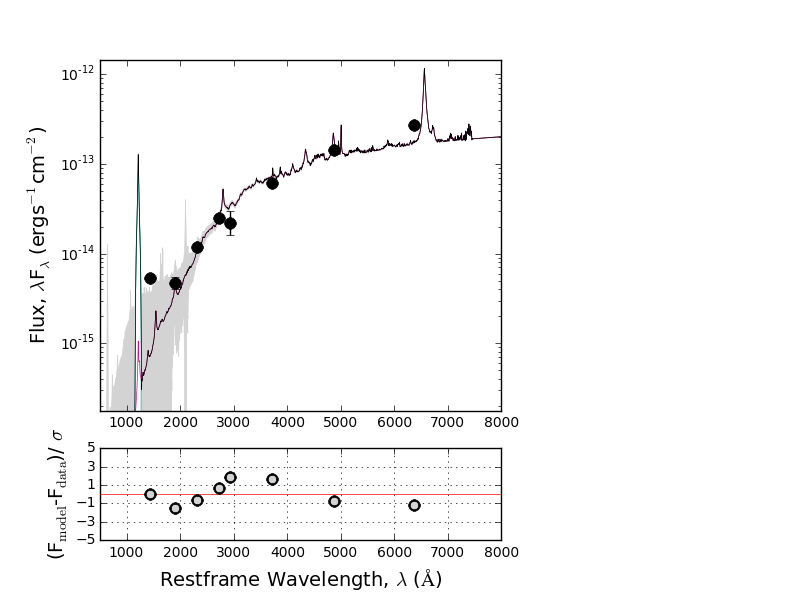}
    \caption{Best-fit model (black) for VHSJ2306-5447 with individual Lyman-$\alpha$ (blue) and reddened quasar (pink) components overlaid.}
    \label{fig:fig_lyman}
\end{figure}

Fitting this model returns $\chi^{2}_{\rm{red,ly\alpha}}$ > 2.5 for four of the six quasars listed above, indicating that scattered Lyman-$\alpha$ emission cannot accurately reproduce the rest-frame UV flux observed in these systems. Although both ULASJ2200+0056 and VHSJ2306-5447 return $\chi^{2}_{\rm{red,ly\alpha}}$ $\sim$2, we note that excess rest-frame UV emission relative to the reddened quasar model is detected in multiple bands in ULASJ2200+0056, further ruling out the scattered Lyman-$\alpha$ scenario for this quasar. From this, we conclude VHSJ2306-5447 to be the only quasar for which the rest-frame UV flux may arise solely from Lyman-$\alpha$ emission. The best fit SED of this scenario for VHSJ2306-5447 in shown in Fig.~\ref{fig:fig_lyman}. For the remaining 12 quasars, we require some other mechanism to characterise the observed emission in the rest-frame UV.

\subsubsection{Scattered Quasar Continuum Light}
\label{sec:scattered_light} 

It is also possible that some fraction of the rest-frame UV quasar continuum light is being scattered into our line of sight by the surrounding dust. The extent of this scattering is assumed to be wavelength-independent and is modelled with two quasar templates, one of which is completely un-obscured to represent the scattered continuum. The dust reddening of the obscured quasar and the fraction of scattered continuum form the two free parameters in the fitting, taking values 0.3 $\leq E(B-V)_{\rm{QSO}} \leq$ 5.0 and 0 $\leq f_{\rm{QSO}} \leq$ 1 respectively, where an $f_{\rm{QSO}}$ = 1 represents the complete scattering of the quasar continuum. This model generally appears to provide a good fit to the photometry, finding a scattering fraction $\sim$0.2 per cent across all sources and returning a $\chi^{2}_{\rm{red,cont}}$ < 2.5 in 10 of the 13 quasars, with only three quasars - VHSJ2028-5740, VHSJ2227-5203 and VHSJ2256-4800 - returning $\chi^{2}_{\rm{red,cont}}$ > 2.5. 

The \textsc{modest} classifier show evidence for spatially extended emission in all but two (ULASJ2200+0056 and ULASJ2224-0015) of the quasars well-fit by this model (Section~\ref{sec:data}). \cite{obied16} show that giant 'scattering cones' of free electrons and dust can scatter the quasar continuum out to $\sim$ 10kpc, resulting in extended regions with narrow line emission and indeed integral field spectroscopy of luminous, obscured quasars have shown narrow-line regions extended over scales of $\sim$20-30 kpc \citep{liu13}. The NIR spectra of our reddened quasar sample \citep{banerji12,banerji15} however, show emission from narrow line regions to be weak, thus ruling out the presence of such a scattering cone. Considering the small fraction of quasar light being scattered in our sample ($\sim$0.2 per cent), it is also possible that extended emission arises from scattering within the interstellar medium (ISM) \citep[e.g.][]{young09}. We note however, that in an ionised ISM with a conventional dust-to-gas ratio, scattering by dust has been shown to dominate over free-electron scattering by a factor $\sim$10 in the UV \citep{weingartner01,draine03}. \cite{draine03} have shown this dust scattering to be strongly biased towards forward scattering, meaning the single scattering of the quasar continuum is unlikely to scatter into the line of sight when considering Type-1 quasars, such as those in our sample. Furthermore, if a small fraction of this emission were to be scattered into our line of sight, we would expect a much bluer quasar spectrum than the fitted unobscured quasar SED, as the scattering efficiency rapidly increases at shorter wavelengths. We find no evidence to suggest a bluer quasar template would better fit the $gri$ DES photometry of our sample and hence conclude this scenario to be unlikely. Although we cannot rule out the scattered light scenario in explaining the UV excess in ULASJ2200+0056 or ULASJ2224-0015, where we see no evidence for extended emission, we likely require an alternative explanation for the remainder of our quasar sample.

The above SED model used to test for the scattered quasar continuum could also represent a system with two AGN (e.g. following a merger event) in which only one is significantly obscured by dust. We quickly rule out the possibility of clustered quasars, as their typical angular separations (2.9 $\leq$ $\theta$ $\leq$ 7.7 arcsec \cite{eftekharzadeh17}) exceed the $\simeq$1\,arcsec angular resolution of DES, yet we see no evidence for a second quasar in the DES images (Fig.~\ref{fig:colour_images}). Thus, if our systems were to contain two quasars they would need to be at small physical separations, but have very different line-of-sight extinctions. Based on our fitting results, the second quasar would also need to have L$_{\rm{bol,QSO}}$ $\sim$ 500 times ($\sim$7 magnitudes) fainter that the primary (reddened) quasar and although there are very few constraints on the number densities of such low-luminosity AGN, the combination of properties required for this explanation to hold makes the scenario somewhat implausible.

\subsection{Host Galaxy Emission}
\label{sec:qso_hosts}

Finally, we explore the possibility of the rest-frame UV flux arising from a relatively un-obscured star-forming host galaxy modelled by the galaxy templates outlined in Section~\ref{sec:sed_modelling}, assuming a constant SFH. Although the real SFH of the galaxy is likely to consist of bursts of star formation, alongside phases of quiescence and steady formation, we later demonstrate this assumption of a constant SFH to provide a good description of the rest-frame UV photometry for the vast majority of our sample.

\begin{figure} 
	\includegraphics[width=0.5\textwidth]{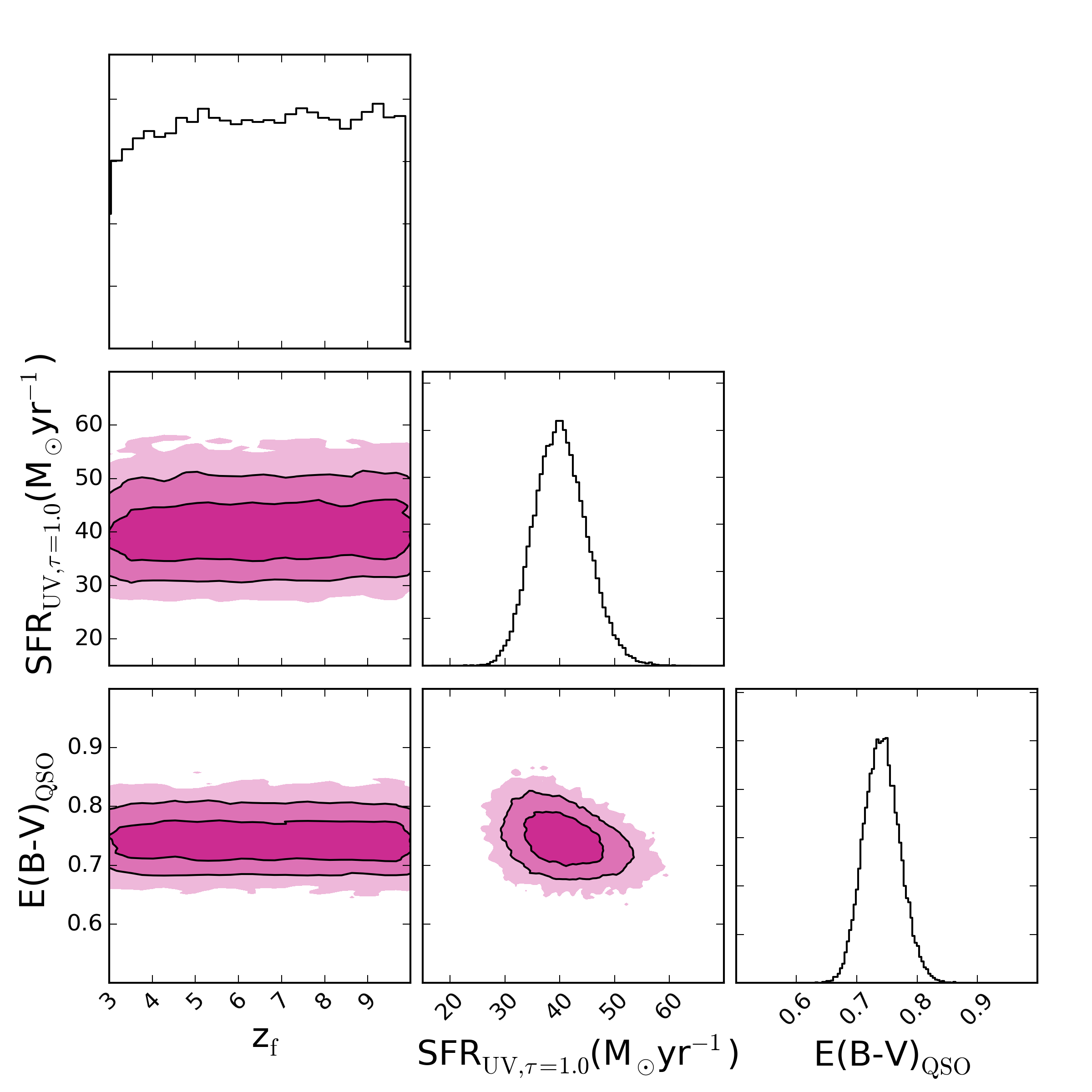}
    \caption{1D and 2D posterior distributions from the MCMC fitting for VHSJ2235-5750 - a representative quasar from the sample. Shaded regions in the 2D distributions denote 1,2 and 3$\sigma$ parameter uncertainties in the fitting. The SFR and associated uncertainties (prior to dust corrections) are based on a galaxy template with $\tau_{\rm{V}}$ = 1.0 and have been converted from the normalisation of the galaxy template ($f_{\rm{gal}}$) in the fitting. Histograms illustrate the relative 1D probability distributions of each parameter.}
    \label{fig:corner}
\end{figure}

Throughout the fitting we consider three free parameters - $E(B-V)_{\rm{QSO}}$, the formation redshift, $z_f$, and the normalisation of the galaxy template, $f_{\rm{gal}}$ - set to take values 0.3 $\leq E(B-V)_{\rm QSO} \leq$ 5.0, 3 $\leq z_{f} \leq$ 10 and 0 $\leq$ $f_{\rm{gal}}$ $\leq$ 1 respectively, where an $f_{\rm gal}$ = 1 denotes a galaxy flux equal to that of the quasar template in the rest-frame UV ($\sim$ 900-3000\AA) prior to any quasar dust reddening. We note that the value of $f_{\rm{gal}}$ is primarily constrained by the DES photometry tracing the rest-frame UV emission and therefore directly traces the UV flux from young stars. Although we are unable to constrain the total mass of the host galaxy, we use the mass of the fitted galaxy template\footnote{Masses we calculated using EzGal (www.baryons.org)}, assuming a constant SFH and a normalisation, $f_{\rm{gal}}$, to derive \emph{instantaneous} SFRs based on the rest-frame UV flux. In this way, $f_{\rm{gal}}$ can be directly converted to an instantaneous SFR (SFR$_{\rm{UV,\tau_{v}=1.0}}$). Figure~\ref{fig:corner} illustrates parameter solutions derived from the MCMC fitting routine for a typical quasar - VHSJ2235-5750 - where this conversion from $f_{\rm{gal}}$ to SFR$_{\rm{UV,\tau_{v}=1.0}}$ has been made (solutions for all 17 quasars can be found in Fig.~\ref{fig:corner20}).

For every quasar in our sample, we recover well-constrained solutions for both $E(B-V)_{\rm QSO}$ and SFR$_{\rm{UV,\tau_{v}=1.0}}$, despite being unable to constrain the age of the system. This indicates that the inferred SFR$_{\rm{UV,\tau_{v}=1.0}}$ is independent from $z_{f}$. We therefore remove $z_{f}$ as a free parameter in the fitting, choosing instead to adopt a fixed value of $z_{f}$ = 6 and solve for the two remaining parameters. The resulting SED fits are presented in the upper panels of Fig.~\ref{fig:seds}. The lower panels of Fig.~\ref{fig:seds} show the residuals of the fit, which are found to be small in all bands, typically returning ($F_{\rm{phot}}-F_{\rm{mod}}$) < 2$\sigma$. For nine of the 13 quasars in Fig.~\ref{fig:seds} we derive $\chi^{2}_{\rm{red,GAL+QSO}}$ < 2.5, indicating the rest-frame UV flux in these objects is well fit by a galaxy template with $\tau_{\rm V}$ = 1.0. Even in the remaining four quasars for which we derive $\chi^{2}_{\rm{red,GAL+QSO}}$ > 2.5, we recover a much better fit than with a reddened quasar template alone (Section~\ref{sec:quasar_only}), finding ($\chi^{2}_{\rm{red,QSO}}$ - $\chi^{2}_{\rm{red,GAL+QSO}}$) > 9 for 11 quasars and ($\chi^{2}_{\rm{red,QSO}}$ - $\chi^{2}_{\rm{red,GAL+QSO}}$) > 3 for the remaining two. This shows that the addition of a host galaxy component to the model significantly improves the goodness of fit for all 13 quasars.

\begin{figure*}
	\centering    
	\subfigure[ULASJ1002+0137]{\label{fig:sed1}\includegraphics[trim=20 -5 200 40 ,clip,width=0.33\textwidth]{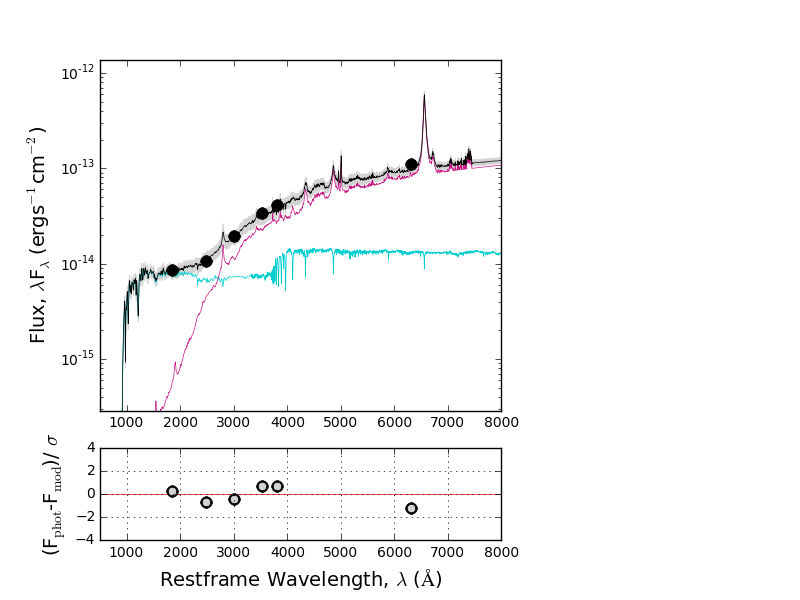}}\hfill
	\subfigure[VHSJ2028-5740]{\label{fig:sed2}\includegraphics[trim=20 -5 200 40 ,clip,width=0.33\textwidth]{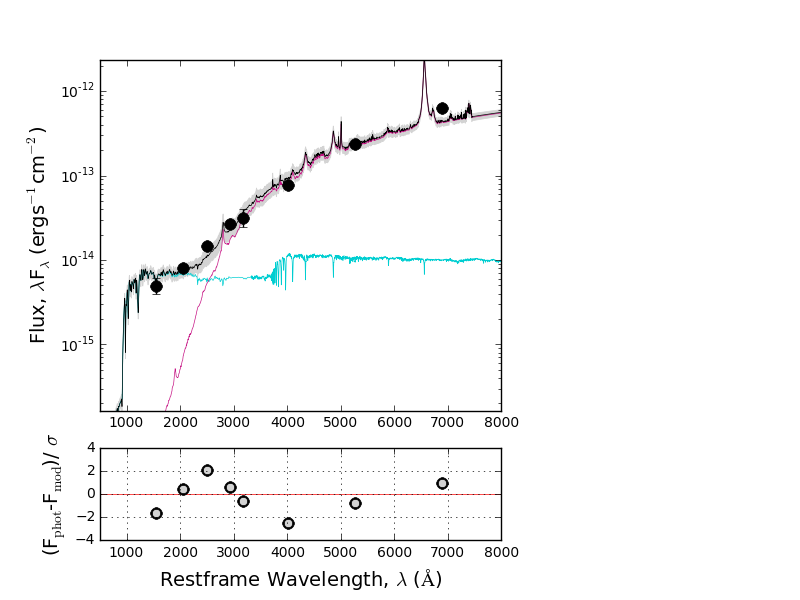}}\hfill
    \subfigure[VHSJ2115-5913]{\label{fig:sed3}\includegraphics[trim=20 -5 200 40 ,clip,width=0.33\textwidth]{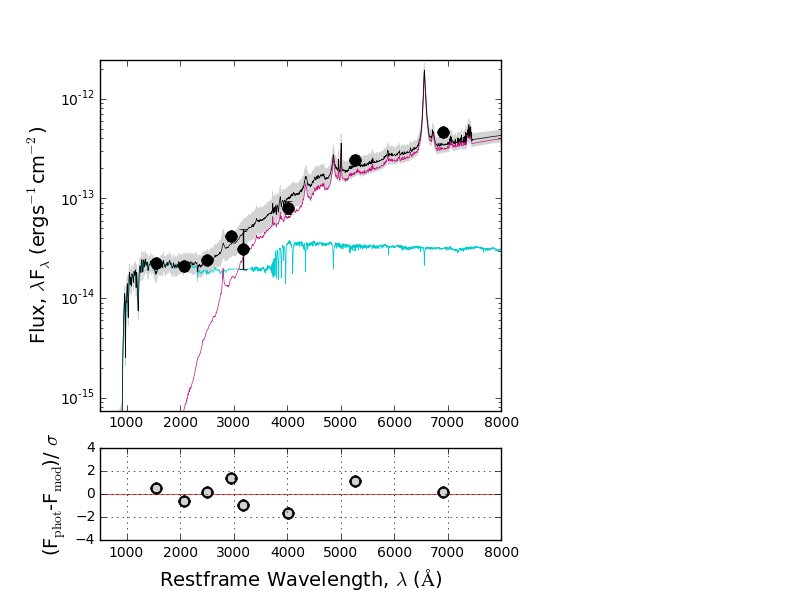}}\hfill
    \subfigure[ULASJ2200+0056]{\label{fig:sed4}\includegraphics[trim=20 -5 200 40 ,clip,width=0.33\textwidth]{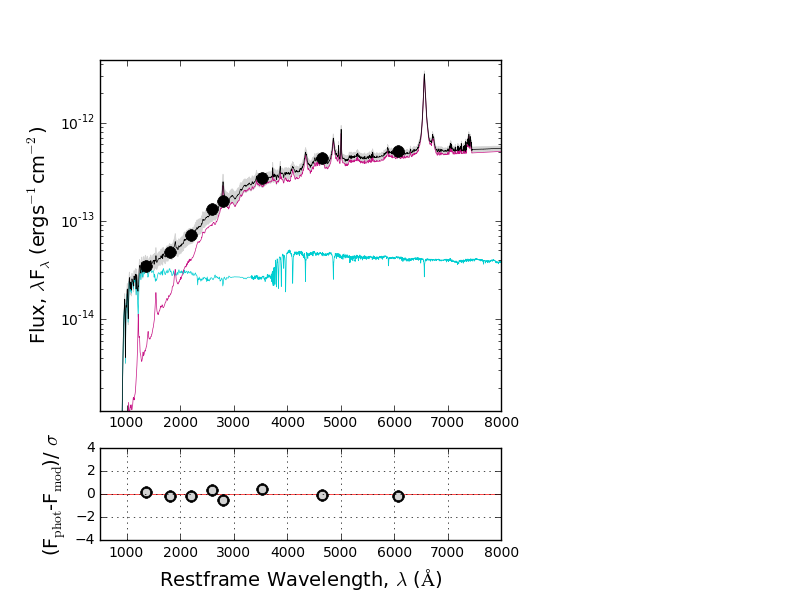}}\hfill
    \subfigure[VHSJ2220-5618]{\label{fig:sed5}\includegraphics[trim=20 -5 200 40 ,clip,width=0.33\textwidth]{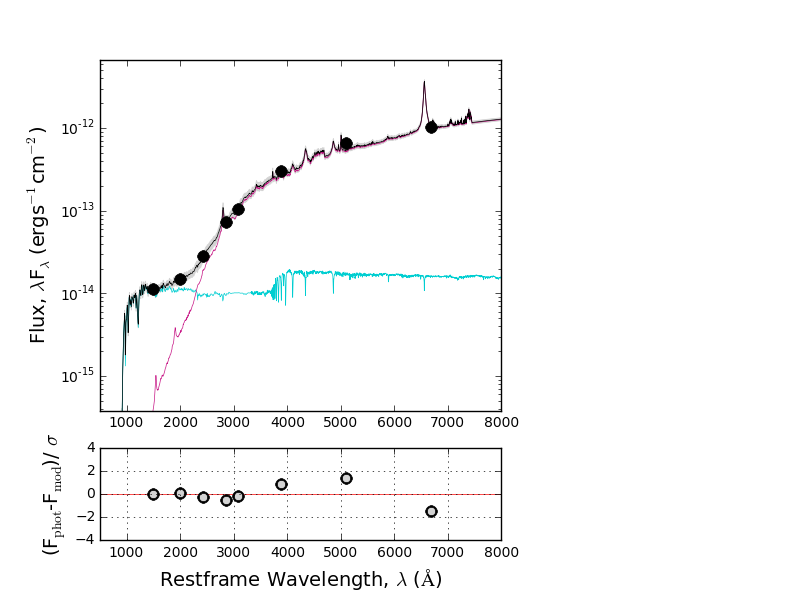}}\hfill
    \subfigure[ULASJ2224-0015]{\label{fig:sed6}\includegraphics[trim=20 -5 200 40 ,clip,width=0.33\textwidth]{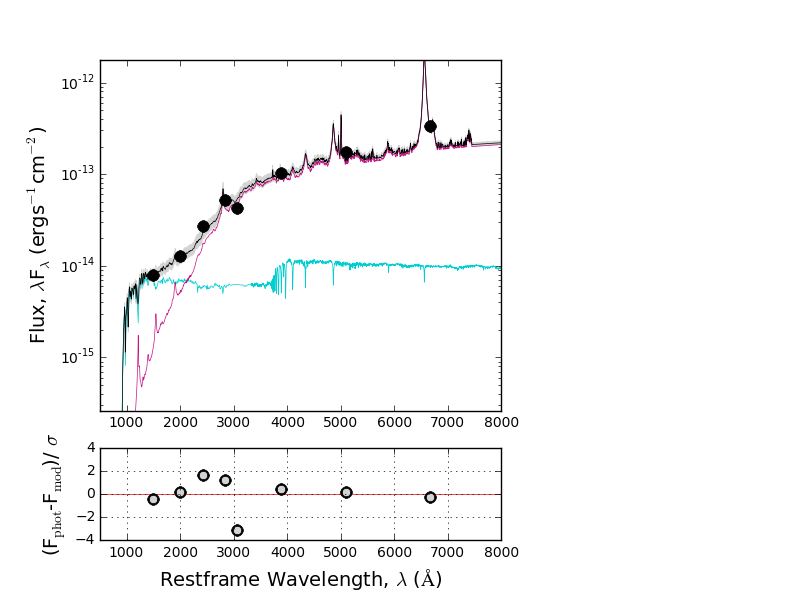}}\hfill
    \subfigure[VHSJ2227-5203]{\label{fig:sed7}\includegraphics[trim=20 -5 200 40 ,clip,width=0.33\textwidth]{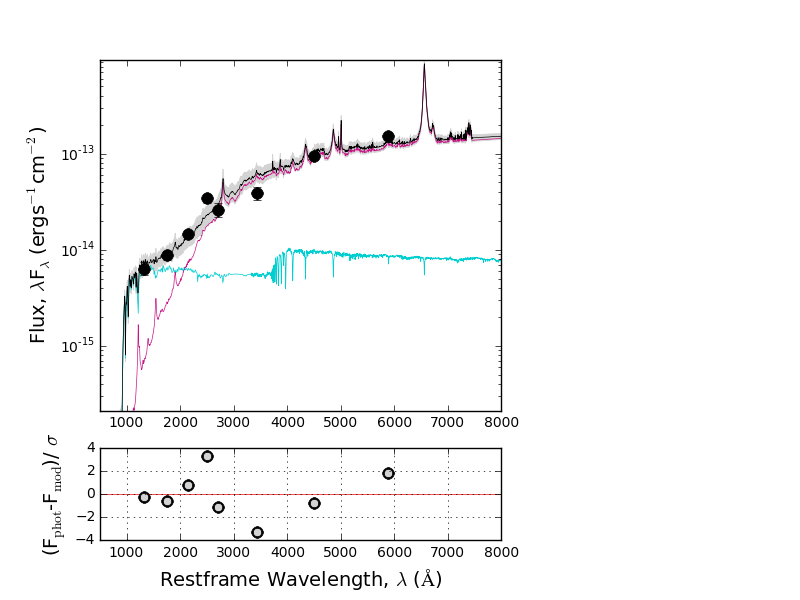}}\hfill
    \subfigure[VHSJ2235-5750]{\label{fig:sed8}\includegraphics[trim=20 -5 200 40 ,clip,width=0.33\textwidth]{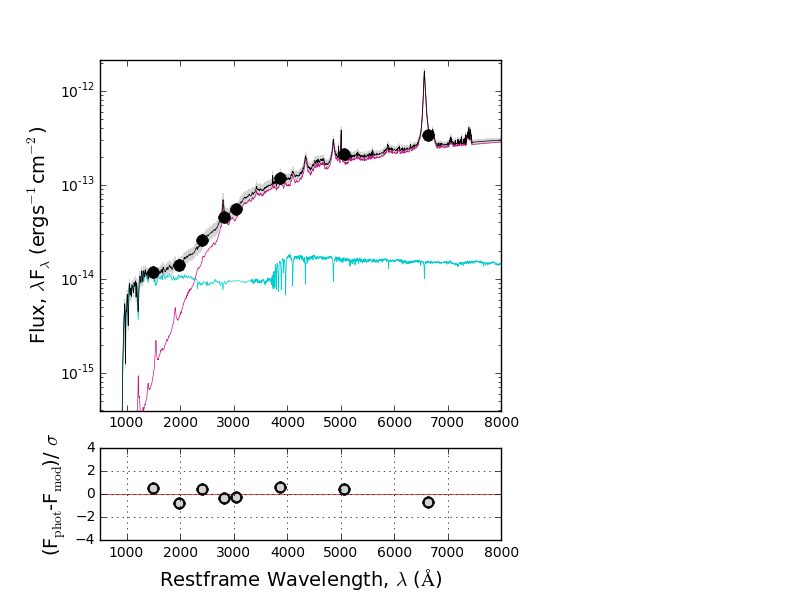}}\hfill
    \subfigure[VHSJ2256-4800]{\label{fig:sed9}\includegraphics[trim=20 -5 200 40 ,clip,width=0.33\textwidth]{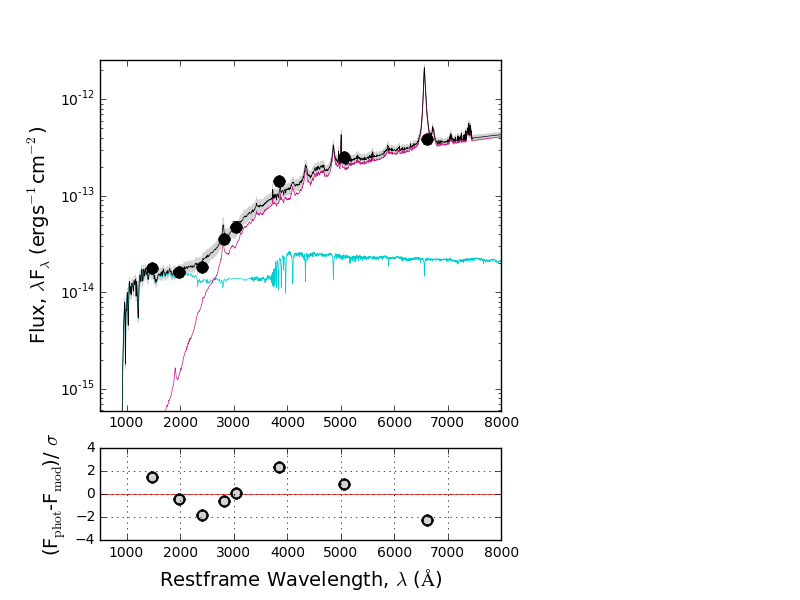}}
\caption{\emph{Upper panels:} Best fit models (black) for the 13 reddened quasars with individual galaxy (blue) and quasar (pink) components overlaid. Black points show the $grizY$ DES and $JHK$ VHS/ULAS photometry. Grey shaded regions illustrate 1$\sigma$ uncertainty about the best-fit model solution. \emph{Lower panels:} Residuals for all photometric bands, scaled to the photometric errors (floored at 10 per cent).}
\label{fig:seds}
\end{figure*}

\begin{figure*}
\addtocounter{figure}{-1}
	\centering
    \subfigure[VHSJ2306-5447]{\label{fig:sed10}\includegraphics[trim=20 -5 200 40 ,clip,width=0.33\textwidth]{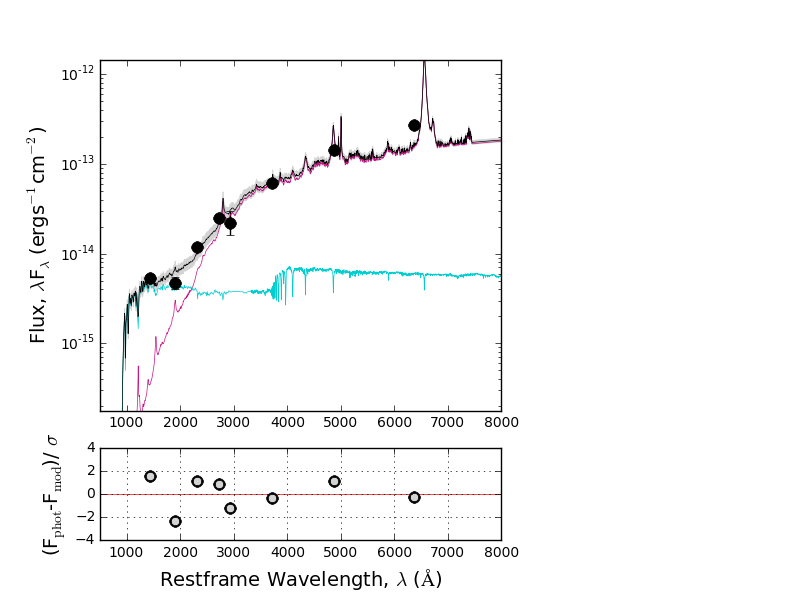}\addtocounter{subfigure}{9}}
    \subfigure[ULASJ2315+0143]{\label{fig:sed11}\includegraphics[trim=20 -5 200 40 ,clip,width=0.33\textwidth]{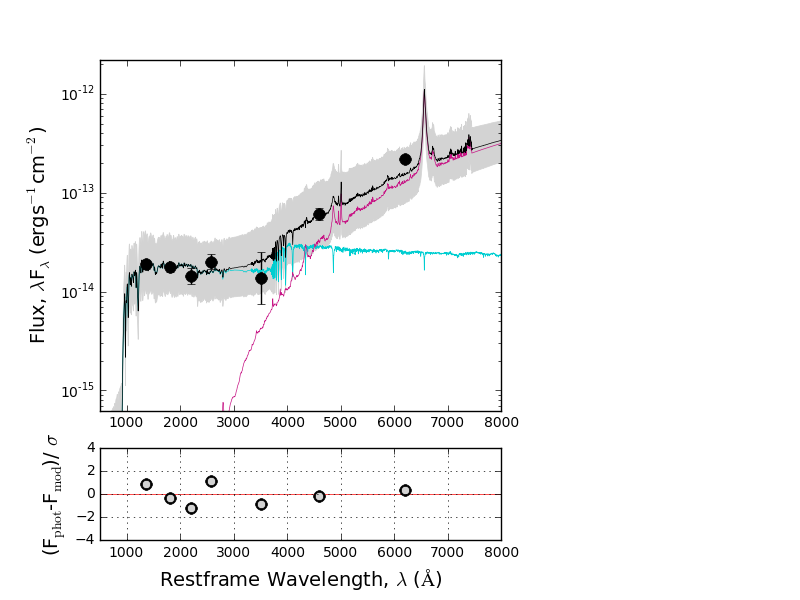}}
    \subfigure[VHSJ2332-5240]{\label{fig:sed12}\includegraphics[trim=20 -5 200 40 ,clip,width=0.33\textwidth]{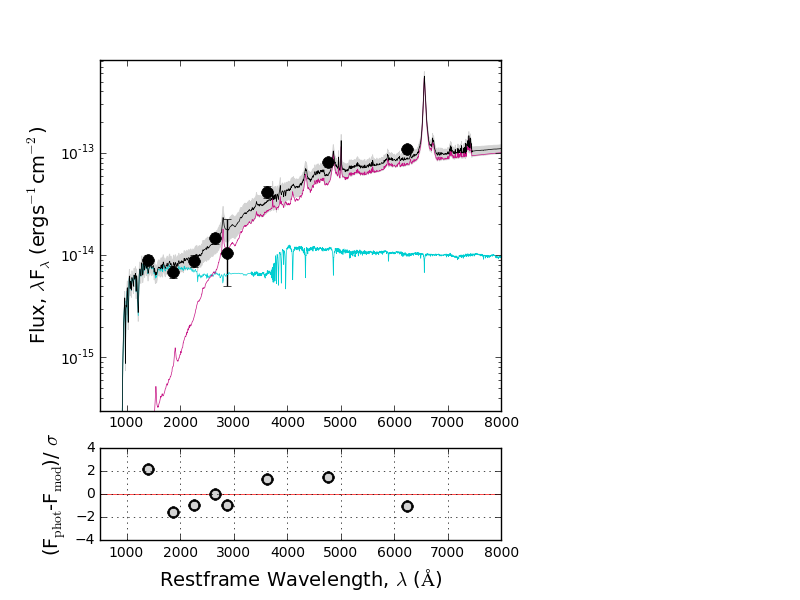}}  
    \begin{center}
    \subfigure[VHSJ2355-0011]{\label{fig:sed13}\includegraphics[trim=20 -5 200 40 ,clip,width=0.33\textwidth]{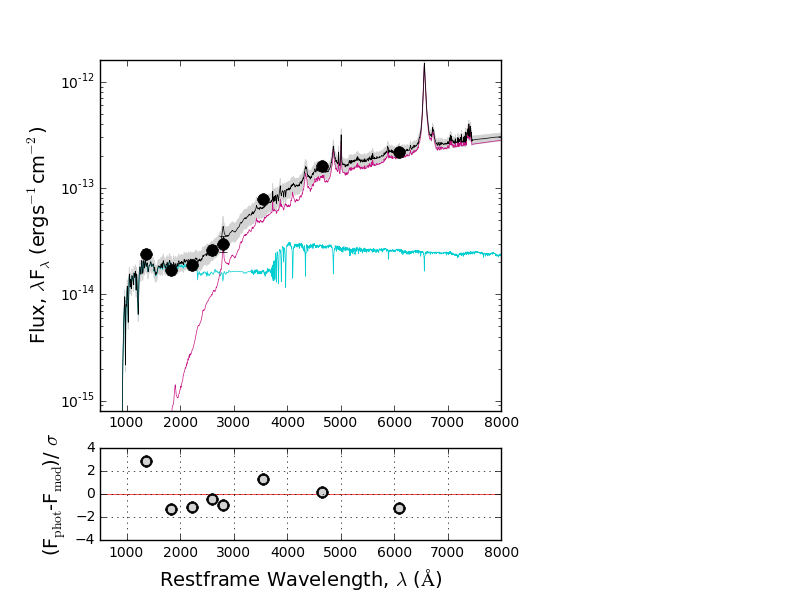}}
    \end{center}
\contcaption{}
\label{fig:seds2}
\end{figure*}

Table~\ref{tab:sfrs} presents instantaneous SFRs derived from the UV flux of the fitted galaxy SEDs in Fig.~\ref{fig:seds}, where we have assumed $\tau_{\rm{V}}=1.0$ and $z_{f}$ = 6.0. SFRs are given both before (SFR$_{\rm{UV,\tau_{v}=1.0}}$) and after (SFR$_{\rm{UV}}$) dust correction has been applied to the galaxy template. We note that although the \citet{charlot00} extinction curve used to correct for the dust is age-dependent, the assumptions made about $z_{f}$ have a minimal effect on the resulting SFRs, changing the SFR by $\pm$ 5-10 per cent for a change in $z_f$ $\pm$ 1.0. These variations lie below the 1$\sigma$ parameter uncertainties derived from the MCMC fitting, also given in Table~\ref{tab:sfrs}, and so we do not discuss them further. For the four quasars that were well-fit by a single reddened-quasar SED (Section~\ref{sec:quasar_only}), we derive upper limits on their SFRs by repeating the above host galaxy fitting for these objects.

\begin{table*}
	\centering
	\caption{UV-derived SFRs prior to dust correction (SFR$_{\rm{UV,\tau_{v}=1.0}}$) and corrected for a $\tau_{\rm{v}}$ = 1.0 (SFR$_{\rm{UV}}$). $E(B-V)_{\rm{QSO}}$ (assuming $z_{\rm{f}}$ = 6) and $E(B-V)_{\rm{QSO,B}}$ \citep{banerji12, banerji15} are also presented. Quoted errors denote the 1$\sigma$ uncertainties on the best-fit model, found by sampling the SEDs fitted during the MCMC routine. Upper limits on the SFR are given to a 68.27 per cent confidence (1$\sigma$).}
	\label{tab:sfrs}
	\begin{tabular}{lcccc}
		\hline
		\hline
		Name & \multicolumn{1}{|p{3cm}|}{\centering SFR$_{\rm{UV,\tau=1.0}}$ \\ (M$_{\odot}$yr$^{-1}$)} & \multicolumn{1}{|p{3cm}|}{\centering SFR$_{\rm{UV}}$ \\ (M$_{\odot}$yr$^{-1}$)} & $E(B-V)_{\rm QSO}$ & $E(B-V)_{\rm QSO,B}$\\
        \hline  
        ULASJ0016-0038 & $\leq$ 25 & $\leq$ 55 & 0.48 $\pm$ 0.03 & 0.5  \\[3pt]
        ULASJ1002+0137 & 10 $\pm$ 5	& 25 $\pm$ 5 & 0.83 $\pm$ 0.04 & 1.0  \\[3pt]
		VHSJ2024-5623 & $\leq$ 10 & $\leq$ 20 & 0.61 $\pm$ 0.03 & 0.6  \\[3pt]
		VHSJ2028-5740 & 25 $\pm$ 5 & 50 $\pm$ 10 & 1.12 $^{+0.04}_{-0.06}$ & 1.2  \\[3pt]
		VHSJ2100-5820 & $\leq$ 10 & $\leq$ 20 & 0.84 $\pm$ 0.03 & 0.8  \\[3pt]
		VHSJ2115-5913 & 70 $^{+15}_{-10}$ & 160 $\pm$ 30 & 1.08 $^{+0.07}_{-0.08}$ & 1.0  \\[3pt]
        ULASJ2200+0056 & 160 $\pm$ 20 & 375 $\pm$ 50 & 0.58 $\pm$ 0.03 & 0.5  \\[3pt]
		VHSJ2220-5618 & 40 $\pm$ 5 & 95 $\pm$ 10 & 0.94 $\pm$ 0.02 & 0.8  \\[3pt]
        ULASJ2224-0015 & 25 $\pm$ 5	& 60 $\pm$ 5 & 0.72 $\pm$ 0.03 & 0.6  \\[3pt]
		VHSJ2227-5203 & 35 $\pm$ 5 & 85 $^{+15}_{-10}$ & 0.63 $\pm$ 0.04 & 0.9  \\[3pt]	
		VHSJ2235-5750 &	40 $\pm$ 5 & 95 $\pm$ 10 & 0.74 $\pm$ 0.03 & 0.6  \\[3pt]
		VHSJ2256-4800 &	60 $\pm$ 5 & 135 $\pm$ 15 & 0.93 $^{+0.03}_{-0.04}$ & 0.6  \\[3pt]
		VHSJ2257-4700 &	$\leq$ 5 & $\leq$ 10 & 0.69 $^{+0.02}_{-0.03}$ & 0.7  \\[3pt]
		VHSJ2306-5447 &	20 $\pm$ 5 & 45 $\pm$ 5 & 0.81 $\pm$ 0.03 & 0.7  \\[3pt]
		ULASJ2315+0143 & 100 $^{+85}_{-50}$	& 230 $^{+200}_{-115}$ & 1.78 $^{+0.24}_{-0.28}$ & 1.1  \\[3pt]
		VHSJ2332-5240 &	35 $\pm$ 5 & 85 $\pm$ $^{+15}_{-10}$ & 0.79 $\pm$ 0.05 & 0.6  \\[3pt]
		VHSJ2355-0011 &	95 $\pm$ 15 & 225 $\pm$ 35 & 0.90 $\pm$ 0.05 & 0.7  \\[3pt]	
		\hline
	\end{tabular}
\end{table*}

We also present the $E(B-V)_{\rm QSO}$ values calculated in the fitting in Table~\ref{tab:sfrs}, together with those from \cite{banerji12,banerji15} ($E(B-V)_{\rm QSO,B}$), who fit the $JHK$ photometry with a reddened quasar model and an Sb galaxy template to model the quasar host. The introduction of the DES photometry in our study allows both the SED and luminosity of the host galaxy to be better constrained. With the addition of this $grizy$ photometry, we expect to derive higher values of $E(B-V)_{\rm QSO}$ than $E(B-V)_{\rm QSO,B}$ for objects in which we see prominent galaxy emission, as this emission is likely to contribute to the $J$-band ($\sim$ 3800\AA\ ) and therefore reduce the relative fraction of quasar flux in this band. We show this to be true for eight of the 13 quasars in the sample, typically finding $E(B-V)_{\rm QSO}$ to be $\sim$ 0.1 - 0.3 larger than $E(B-V)_{\rm QSO,B}$. The reddest quasar in our sample - ULASJ2315+0143 - returns an $E(B-V)_{\rm QSO}$ $\sim$ 0.7 larger than $E(B-V)_{\rm QSO,B}$, although the uncertainty on this value is much larger than for the rest of the sample. A further three quasars - VHSJ-2028-5740, VHSJ2115-5913 and ULASJ2200+0056 - return an $E(B-V)_{\rm QSO}$ consistent with $E(B-V)_{\rm QSO,B}$ to within 0.1, with the remaining two quasars - ULASJ1002+0137 and VHSJ2227-5203 - returning an $E(B-V)_{\rm QSO}$ $\sim$ 0.2 - 0.3 lower than $E(B-V)_{\rm QSO,B}$. We note that the $J$-band photometry is not available for ULASJ1002+0137 (Fig.~\ref{fig:sed1}), meaning $E(B-V)_{\rm QSO,B}$ has been constrained from only the $H$ and $K$-band photometry. From Fig.~\ref{fig:sed1}, we find that these two bands indicate a steeper rest-frame NIR slope than is implied when the DES photometry is included in the fit, meaning a model based on just the $H$ and $K$-band data would likely overestimate the quasar reddening. Similarly, the $J$-band photometry in VHSJ2227-5203 lies below our fitted SED model (Fig.~\ref{fig:sed7}), indicating that a higher reddening would be derived without the inclusion of the $grizY$ photometry. We therefore find no direct conflict between the results of \cite{banerji12,banerji15} and the results of this paper.

\subsubsection{UV Slope of the Galaxy SED}
\label{sec:uv_slope} 

\begin{figure}
\centering
    \includegraphics[width=0.5\textwidth]{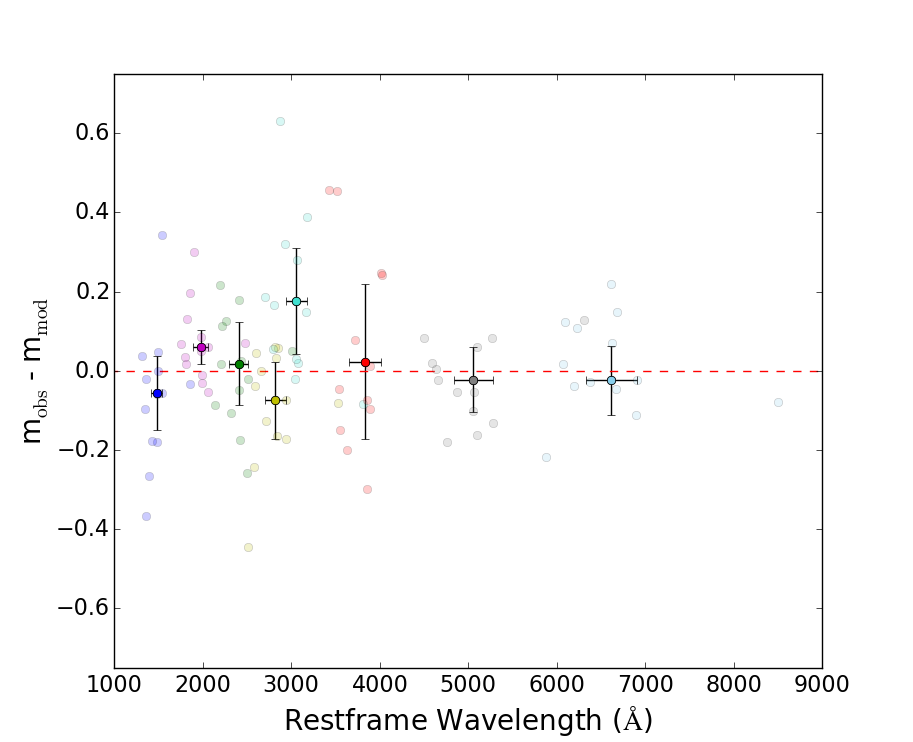}
    \caption{Magnitude differences between the fitted SED and the DES $grizY$ photometry, assuming a galaxy template with $\tau_{\rm{V}}$ = 1.0. Dark points denote the median average in each filter with error bars showing the median absolute deviation of the data points in each passband.}
    \label{fig:delta_mag}
\end{figure}

In general, the galaxy model with a constant-SFH and $\tau_{\rm{V}}$ = 1.0 provides a good fit to our reddened quasar sample. This is demonstrated in Fig.~\ref{fig:delta_mag}, where we present the median average fitting residuals ($m_{\rm data}$ - $m_{\rm model}$) for each band. Four quasars however - VHSJ2028-5740, VHSJ2227-5203, VHSJ2256-4800 and VHSJ2355-0011 - return $\chi^{2}_{\rm{red,GAL+QSO}}$ > 2.5 when this default host galaxy model is assumed. We consider VHSJ2227-5203 further in Section~\ref{sec:qso_variability}, but for the remaining three objects we find the $g$-band photometry to lie below the fitted SED in VHSJ2028-5740 (Fig.~\ref{fig:sed2}) and above the fitted SED in VHSJ2256-4800 (Fig.~\ref{fig:sed9}) and VHSJ2355-0011 (Fig.~\ref{fig:sed13}), indicating that a galaxy model with $\tau_{\rm{V}}$ = 1.0 cannot reproduce the slope of the rest-frame UV flux in these systems. We find that significantly improved $\chi^{2}_{\rm{red,GAL+QSO}}$ values ($\sim$1-2) can be obtained by adopting a galaxy model with $\tau_{\rm{V}}$ = 5.0 for VHSJ2028-5740 and $\tau_{\rm{V}}$ = 0.2 for VHSJ2256-4800 and VHSJ2355-0011. We note however, that the effect of this dust extinction on the rest-frame UV slope is highly degenerate with that caused by changes in the SFH, meaning a similar improvement in $\chi^{2}_{\rm{red,GAL+QSO}}$ may also be obtained by adjusting the SFH. The exact SFRs derived from the SED-fitting are therefore dependent on both the adopted SFH and dust content of the host galaxies, the constraints of which lie beyond the scope of this work given the limited DES photometry. Nevertheless, the detection of rest-frame UV emission consistent with star formation in the sample is robust.

\subsubsection{Quasar Variability}
\label{sec:qso_variability} 

In addition to the assumed SFH and dust content of the host galaxy, quasar variability may also contribute to the large $\chi^{2}_{\rm{red,GAL+QSO}}$ values. The observations from DES and VHS/ULAS were taken about three years apart, corresponding to approximately a year in the rest-frame of our quasars. Over such time scales, luminous quasars have been shown to vary by $\lesssim$ 0.1\,mag in the rest-frame optical \citep[e.g.][]{berk04,de05,macleod12}. Even accounting for the small amplitude of this variability, systematic offsets between the DES and VHS/ULAS photometry are not evident for the sample as a whole. We consider the possibility however, that individual quasars in our sample may exhibit variability with larger amplitudes and for VHSJ2227-5203 ($\chi^{2}_{\rm{red,GAL+QSO}}$ > 2.5), we find the $JHK$-band photometry to appear several tenths of a magnitude fainter than that extrapolated from the $grizY$ photometry. To test whether variability could explain the poor fit of VHSJ2227-5203, the fitting routine is re-run, having increased the $JHK$ magnitudes by 0.2\,mag - the magnitude offset required to align the DES and VHS photometry. However, the fit of this adjusted photometry returned $\chi^{2}_{\rm{red,GAL+QSO}}\simeq 5$, meaning that an offset between the DES and VHS photometry due to quasar variability, cannot account for the large $\chi^{2}_{\rm{red,GAL+QSO}}$ in VHSJ2227-5203. We therefore conclude that photometric quasar variability does not play a significant role in our sample of reddened quasars and therefore does not affect the conclusions of this paper.

\section{Discussion}
\label{sec:discussion}

We have found evidence for an excess of rest-frame UV emission, relative to a reddened quasar SED, in 13 of the 17 quasars considered in this paper. For each quasar in our sample, we have explored three possible sources of this emission - resonantly scattered Lyman-$\alpha$ emission, scattered light from the quasar continuum and a relatively un-obscured star forming host, demonstrating the rest-frame UV emission to be consistent with star formation in at least 11 quasars. Furthermore, we have found two quasars - VHSJ2306-5447 and ULASJ2224-0015 - to exhibit rest-frame UV emission consistent with either star formation or scattered light and have derived instantaneous SFRs across the entire quasar sample, based upon their rest-frame UV flux. The following section compares our results to a range of independent studies and outlines connections between the properties of these luminous, high-redshift, reddened quasars and their host galaxies.

\subsection{Comparison to other SFR Estimates for Heavily Reddened Quasars}
\label{sec:comparisons}

\subsubsection{H$\alpha$ SFRs}
\label{sec:h_alpha}

\begin{figure}
	\includegraphics[width=0.5\textwidth]{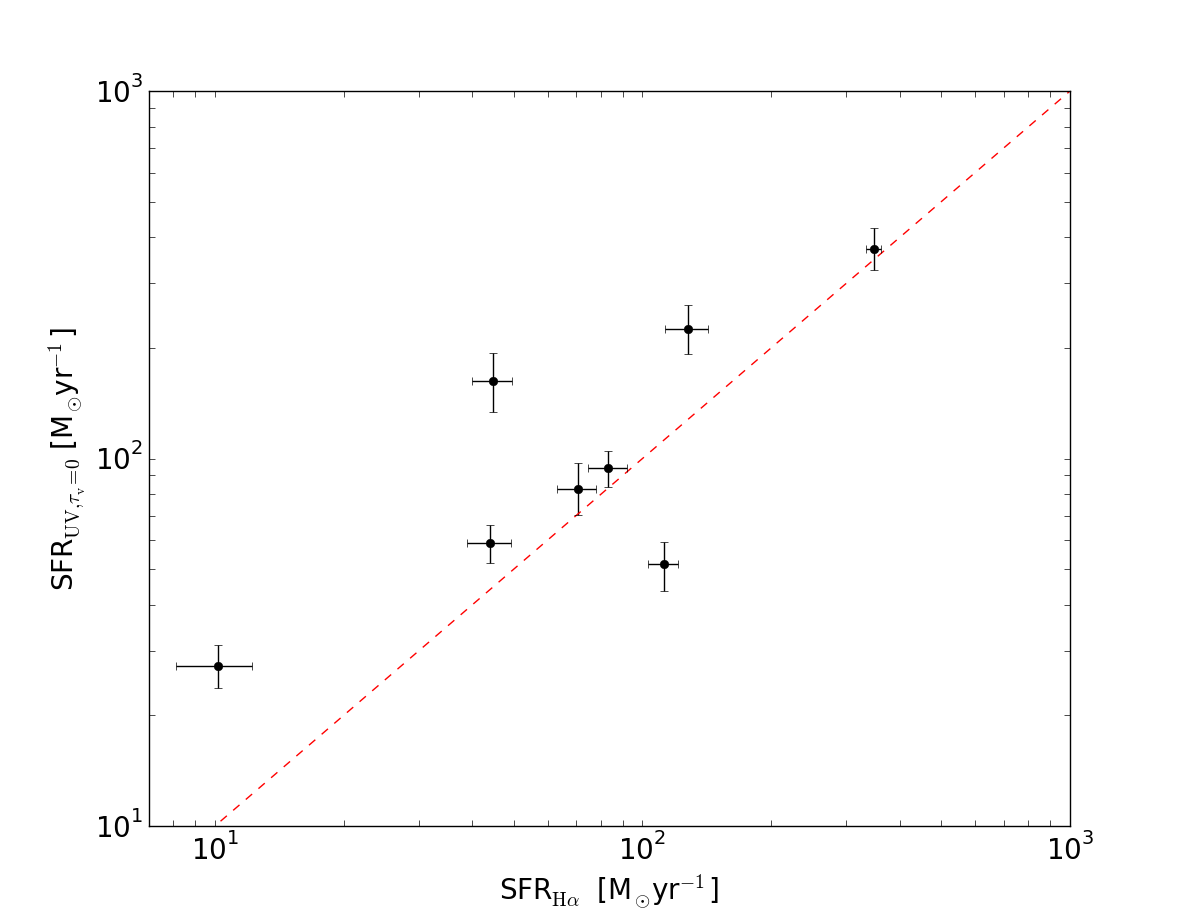}
    \caption{Comparison of SFR$_{\rm{UV}}$ to SFR$_{\rm{H\alpha}}$ \citep{alaghband2016}. Error bars show the 1$\sigma$ uncertainties on the calculated SFRs in each case.} 
    \label{fig:h_alpha}
\end{figure}

Fifteen of the reddened quasars in our sample overlap with those in \cite{alaghband2016}, who look for narrow H$\alpha$ emission that could be associated with star formation in the NIR spectra. Narrow H$\alpha$ emission is detected in nine of the 15 quasars in their sample, from which \cite{alaghband2016} derive extinction-corrected SFRs assuming a Calzetti dust extinction law \citep{calzetti00} and a dust attenuation towards the star-forming regions equal to that towards the quasar continuum. In our analysis, one of these nine quasars - VHSJ2100-5820 - does not show an excess of rest-frame UV emission relative to that expected from a reddened quasar and so is not considered further in our comparisons to H$\alpha$-derived SFRs. Similarly, we do not include the six quasars with no evidence of narrow H$\alpha$ emission, for which \citet{alaghband2016} derive only upper limits on the SFR, assuming the star-forming regions to be located within the PSF of the SINFONI data (FWHM $\sim$0.5-0.8\,arcsec). As we clearly see evidence for rest-frame UV emission on larger scales than this across our sample, the H$\alpha$ SFR limits for these six quasars are not directly comparable to our work. We therefore only compare our UV-derived SFRs with the H$\alpha$ SFRs for the eight quasars outlined in Table~\ref{tab:halpha_sfrs}, where the H$\alpha$ SFRs have been re-derived, correcting for $\tau_{\rm{V}}$ = 1.0, as has been assumed for our galaxy template. Fig.~\ref{fig:h_alpha} shows the comparison between these two measures of SFR.

\begin{table}
	\centering
	\caption{SFR$_{\rm{H\alpha}}$ for the eight quasars overlapping our host galaxy sample, for which \citet{alaghband2016} detect narrow H${\alpha}$ emission. SFRs have been corrected for dust, assuming an $E(B-V)$ = 0.35 to be consistent with the level of dust used in our galaxy template ($\tau_{\rm{V}}\sim$1.0).}
	\label{tab:halpha_sfrs}
	\begin{tabular}{lc}
		\hline
		\hline
		Name & SFR$_{\rm{H\alpha}}$ (M$_{\odot}$yr$^{-1}$)\\
        \hline  
        ULASJ1002+0137 & 10 $\pm 5$ \\[3pt]
		VHSJ2028-5740 & 110 $\pm 10$ \\[3pt]
		VHSJ2115-5913 & 45 $\pm 5$ \\[3pt]
        ULASJ2200+0056 & 350 $\pm 15$ \\[3pt]
        ULASJ2224-0015 & 45 $\pm 5$ \\[3pt]
		VHSJ2235-5750 & 85 $\pm 10$ \\[3pt]
		VHSJ2332-5240 & 70 $\pm 10$ \\[3pt]
		VHSJ2355-0011 & 130 $\pm 15$ \\[3pt]	
		\hline
	\end{tabular}
\end{table}

Fig.~\ref{fig:h_alpha} indicates the two measures of star formation to be broadly correlated, although values of SFR$_{\rm{UV}}$ appear systematically higher than SFR$_{\rm{H\alpha}}$ by a factor of $\sim$1.1-4. This lack of a one-to-one correspondence between the two SFR measures is expected however, as several assumptions have been made regarding the quasar host galaxy and the dust attenuation affecting both the UV- and H$\alpha$-emitting regions. We also note that \citet{alaghband2016} restrict their search for narrow H$\alpha$ emission to $\sim$0.5-0.8\,arcsec regions (corresponding to the size of the PSF in their observations), meaning their study is not sensitive to extended emission on larger scales. Given that the rest-frame UV emission in our sample typically extends $\gtrsim$1 arcsec, we expect the SFRs we derive from the spatially-integrated UV flux to generally be higher than those measured from H$\alpha$ by \citet{alaghband2016}, which is exactly what we see in Fig.~\ref{fig:h_alpha}

\subsubsection{\textit{Herschel} and ALMA SFRs}
\label{sec:alma}
Several of the reddened quasars in our sample have also been detected in the far infra-red and millimetre wavelengths, suggesting significant levels of dust-obscured star formation in these sources \citep{banerji14,banerji17} exceeding 1000M$_\odot$ yr$^{-1}$. Recently, \citet{banerji17} studied four reddened quasars with ALMA, finding ubiquitous evidence for large reservoirs of molecular gas and star formation in all four systems. One of these four quasars - ULASJ2315+0143 - overlaps with our sample, and shows resolved CO(3-2) emission elongated in the same direction as the rest-frame UV emission seen in Fig.~\ref{fig:colour16}. The ALMA-derived SFR for this source is 6100$\pm$400 M$_\odot$yr$^{-1}$ and whilst this is significantly larger than the 230M$_\odot$yr$^{-1}$ UV-derived SFR inferred from the DES observations, we note that the ALMA SFR is derived from a single photometric data point at $\nu$=103GHz tracing the Rayleigh-Jeans tail of the dust SED, and could be over-estimated for a number of reasons. Specifically, \citet{banerji17} assume no significant contribution from radio synchrotron emission to the millimeter continuum emission, and the dust heating is assumed to be entirely dominated by the star-forming host, without any contribution from the quasar itself. Nevertheless, given the significant amount of dust emission seen by ALMA, it is plausible that the rest-frame UV emission in the DES $grizY$ bands is only tracing a small fraction of the total star formation in this galaxy.

\subsection{Comparison to Hot Dust-Obscured Galaxies (HotDOGs)}

Hot Dust Obscured Galaxies (HotDOGs) are a related class of heavily obscured AGN selected using data from the Wide-Field Infra-red Survey Explorer (WISE). They have similar intrinsic AGN luminosities to our reddened quasars but much higher levels of dust obscuration towards the quasar sight-line \citep{eisenhardt12}. \citet{assef16} recently presented evidence for excess rest-frame UV emission in several HotDOGs and conducted a detailed multi-wavelength investigation of one particular source - W0204-0506 - at $z=2.1$. In their investigation, \citet{assef16} consider scattered AGN light, a second AGN and a star-forming host galaxy as possible sources of the UV emission seen in W0204-0506, concluding scattered light from the AGN to be the most plausible scenario. One reason \citet{assef16} give for this conclusion is the unprecedented level of un-obscured star formation required to reproduce the UV flux observed in W0204-0506.

To compare W0204-0506 with the reddened quasars in our sample, we fit the photometry for this source with our constant SFH host galaxy template using the SED- fitting method outlined in Section~\ref{sec:sed_modelling}. From this, we derive a dust-corrected SFR$\sim$5200M$_{\odot}$yr$^{-1}$ for W0204-0506, consistent with the $\sim$5500M$_{\odot}$yr$^{-1}$ derived by \cite{assef16}. This SFR > 5000M$_{\odot}$yr$^{-1}$ is extreme compared to the moderate SFRs we derive from the rest-frame UV flux of our reddened quasars ($\sim$130M$_{\odot}$yr$^{-1}$), although as discussed in Section \ref{sec:alma}, this does not preclude a much higher level of obscured star formation in our sources. Nevertheless, the blue excess \textit{WISE} HotDOGs appear clearly distinct in terms of their properties from the reddened quasar sample considered in this work and hence likely have different mechanisms responsible for the emission in the rest-frame UV.

\subsection{Is SFR correlated with Quasar Luminosity?}
\label{sec:quasar_Lbol}

\begin{figure}
	\includegraphics[width=0.5\textwidth]{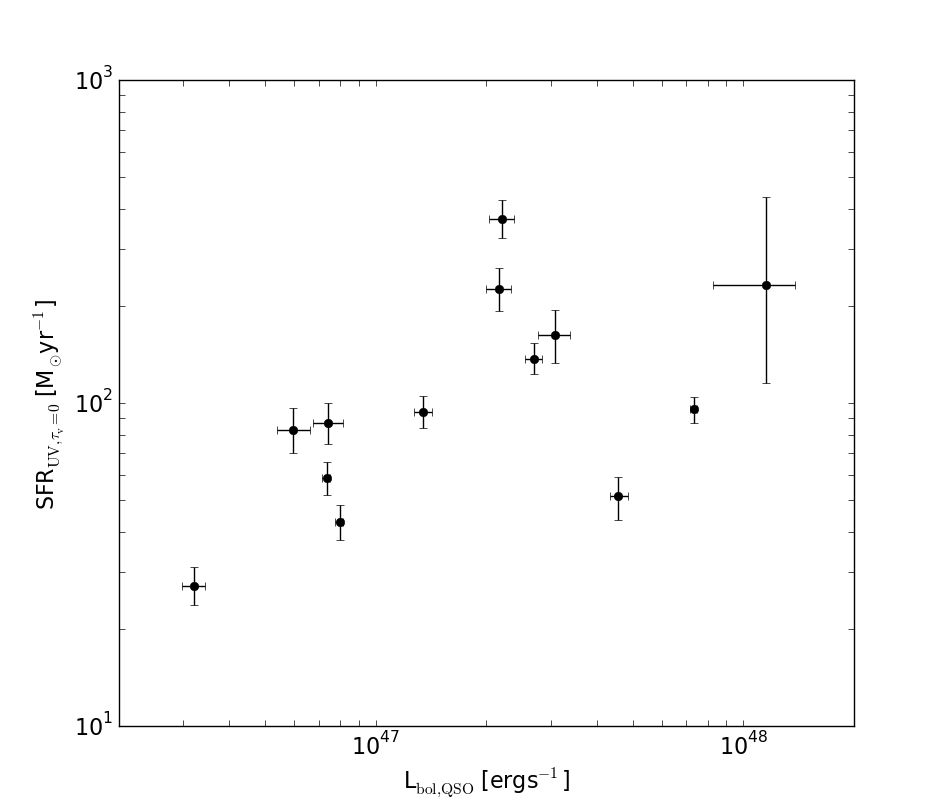}
    \caption{SFR$_{\rm{UV}}$ as a function of the dust-corrected L$_{\rm{bol,QSO}}$. Error bars denote 1$\sigma$ uncertainties on SFR$_{\rm{UV}}$.} 
    \label{fig:l_bol}
\end{figure}

We now explore whether the instantaneous SFRs we derive for the reddened quasar sample are correlated with the bolometric quasar luminosity, L$_{\rm{bol,QSO}}$. L$_{\rm{bol,QSO}}$ is calculated from the extinction-corrected rest-frame flux at 5100\AA, using a bolometric correction factor of seven \citep{vestergaard06}. The values of L$_{\rm{bol,QSO}}$ we derive are consistent with those presented in \cite{banerji12,banerji15}, with only small variations due to the inclusion of the DES photometry in our fitting. SFR$_{\rm{UV}}$ is plotted as a function of L$_{\rm{bol,QSO}}$ in Fig.~\ref{fig:l_bol}, where we see a positive correlation between these two quantities. From Fig.~\ref{fig:l_bol} we find a Spearman's correlation coefficient of r $\simeq$ 0.6 with a p-value, $\alpha$ = 0.03, i.e. we detect a positive correlation to a 97 per cent confidence level. Although SFR$_{\rm{UV}}$ and L$_{\rm{bol,QSO}}$ have both been corrected for dust extinction, we note that the apparent correlation in Fig.~\ref{fig:l_bol} is not due to this correction, as different corrections have been applied to the quasar and the galaxy individually (E(B-V)$_{\rm{GAL}}$=0.35, while E(B-V)$_{\rm{QSO}}$ $\sim$0.5-2.0, as solved for in the fitting). We also note that the observed correlation in Fig.~\ref{fig:l_bol} is not caused by the simultaneous fitting of the galaxy and quasar components in the model, as each of these components is constrained by different photometry. The galaxy SED for example, is primarily constrained by the $gri$ DES photometry, whilst the $JHK$ VISTA/ ULAS photometry is used to constrain the quasar SED, meaning the two components are essentially decoupled. 

The correlation we observe between SFR$_{\rm{UV}}$ and L$_{\rm{bol,QSO}}$ in our reddened quasars (Fig.~\ref{fig:l_bol}) is consistent with the work of \cite{harris16}, who find the SFRs of type-1 quasar host galaxies at $2 < z < 3$ to increase with AGN luminosity up to some limiting rate. \cite{shao10} and \cite{netzer09} find a similar trend among high luminosity AGN (L$_{\rm{bol}}$ > 10$^{44}$ergs$^{-1}$), but find no such dependency in lower luminosity systems - a result confirmed by independent studies of moderate-luminosity AGN \citep[e.g.][]{jahnke04}. Furthermore, \cite{urrutia12} find no correlation between SFR and AGN accretion among reddened quasars at low redshift ($0.4 < z < 1$), suggesting that the trend in Fig.~\ref{fig:l_bol} may only appear for the most luminous quasars at high redshift.

Although we find some evidence for a relationship between L$_{\rm{bol,QSO}}$ and SFR$_{\rm{UV}}$ in our sample, we highlight that this paper only accounts for the un-obscured, instantaneous star formation rate, based upon the rest-frame UV emission observed in a relatively small sample of 13 reddened quasars and that a number of assumptions have been made regarding the host galaxy properties. Assembling larger samples of such measurements as well as measuring the total, rather than un-obscured, SFRs from SCUBA-2 and ALMA observations will help corroborate whether there is indeed a correlation between star formation rate and quasar luminosity in reddened quasars.

\section{Conclusions}
\label{sec:conclusions}

We have presented the first rest-frame UV observations for a population of 17 luminous, heavily dust-reddened ( L$_{\rm{bol,QSO}} > $ 10$^{46}$ergs$^{-1}$, $E(B-V)_{\rm{QSO}} \sim$ 0.8) broad-line quasars at 1.5 < z < 2.7, making use of DES Year 1 optical photometry along with NIR observations from VHS and ULAS. Our main conclusions can be summarised as follows:

i) We detected an excess of rest-frame UV emission relative to that expected from a reddened quasar template alone in 13 of the 17 sources in our sample. Via SED fitting, we modelled three potential sources of this excess rest-frame UV emission in the reddened quasar sample - resonantly-scattered Lyman-$\alpha$, scattered light from the quasar continuum and a star-forming host galaxy. In at least ten quasars, we found the rest-frame UV emission to be consistent with the detection of a relatively un-obscured, star forming host galaxy, with a further three quasars showing emission that could be consistent with either star formation or scattered light. 

ii) We estimated the instantaneous SFRs for the quasar host galaxies in the sample based on their rest-frame UV emission, deriving extinction-corrected star formation rates of 25 $\lesssim$ SFR$_{\rm{UV}}$ $\lesssim$ 365 M$_{\odot}$yr$^{-1}$, with a mean average SFR$_{\rm{UV}}$ = 130 $\pm$ 95 M$_{\odot}$yr$^{-1}$. We note, however, that the SFRs presented in this paper account only for the relatively unobscured component of star formation and therefore do not preclude much higher rates of obscured star formation in these systems. In fact, we see evidence for a much higher obscured SFR in ULASJ2315+0143 based on recent ALMA observations and cannot rule out a similar trend in the remainder of our sample. The unobscured SFRs presented in this paper are found to be broadly consistent with those derived independently from H$\alpha$ observations using Integral Field Spectroscopy.

iii) We have found evidence for a correlation between the rest-frame UV-derived, dust-corrected star formation rates and the bolometric luminosity of the quasar at the high quasar luminosities probed by our sample. This is consistent with similar trends observed for high-luminosity populations of unobscured quasars.  

Overall, we have demonstrated that quasar emission can be disentangled from its host galaxy, even in the most luminous systems, by exploiting dust obscuration towards the quasar. For the first time, populations of these intrinsically-luminous, dusty quasars can now be studied in the rest-frame UV at high redshift thanks to deep, wide-field observations from large imaging surveys like DES. Our analysis has shown that several of the hosts of these reddened quasars are spatially resolved in the DES imaging tracing rest-frame UV wavelengths. This paves the way for future high-resolution imaging and detailed morphological studies of these quasar host galaxies.

\section*{Acknowledgements}
\label{sec:acknowledgements}

CFW thanks the Science and Technology Facilities Council (STFC) for the award of a studentship. MB acknowledges support from STFC via an Ernest Rutherford Fellowship. PCH acknowledges support from the STFC via a Consolidated Grant to the Institute of Astronomy, Cambridge. 

Funding for the DES Projects has been provided by the U.S. Department of Energy, the U.S. National Science Foundation, the Ministry of Science and Education of Spain, the Science and Technology Facilities Council of the United Kingdom, the Higher Education Funding Council for England, the National Center for Supercomputing Applications at the University of Illinois at Urbana-Champaign, the Kavli Institute of Cosmological Physics at the University of Chicago, the Center for Cosmology and Astro-Particle Physics at the Ohio State University,the Mitchell Institute for Fundamental Physics and Astronomy at Texas A\&M University, Financiadora de Estudos e Projetos, Funda{\c c}{\~a}o Carlos Chagas Filho de Amparo {\`a} Pesquisa do Estado do Rio de Janeiro, Conselho Nacional de Desenvolvimento Cient{\'i}fico e Tecnol{\'o}gico and the Minist{\'e}rio da Ci{\^e}ncia, Tecnologia e Inova{\c c}{\~a}o, the Deutsche Forschungsgemeinschaft and the Collaborating Institutions in the Dark Energy Survey. 

The Collaborating Institutions are Argonne National Laboratory, the University of California at Santa Cruz, the University of Cambridge, Centro de Investigaciones Energ{\'e}ticas, Medioambientales y Tecnol{\'o}gicas-Madrid, the University of Chicago, University College London, the DES-Brazil Consortium, the University of Edinburgh, the Eidgen{\"o}ssische Technische Hochschule (ETH) Z{\"u}rich, Fermi National Accelerator Laboratory, the University of Illinois at Urbana-Champaign, the Institut de Ci{\`e}ncies de l'Espai (IEEC/CSIC), the Institut de F{\'i}sica d'Altes Energies, Lawrence Berkeley National Laboratory, the Ludwig-Maximilians Universit{\"a}t M{\"u}nchen and the associated Excellence Cluster Universe, the University of Michigan, the National Optical Astronomy Observatory, the University of Nottingham, The Ohio State University, the University of Pennsylvania, the University of Portsmouth, SLAC National Accelerator Laboratory, Stanford University, the University of Sussex, Texas A\&M University, and the OzDES Membership Consortium.

The DES data management system is supported by the National Science Foundation under Grant Number AST-1138766.
The DES participants from Spanish institutions are partially supported by MINECO under grants AYA2015-71825, ESP2015-88861, FPA2015-68048, SEV-2012-0234, SEV-2012-0249, and MDM-2015-0509, some of which include ERDF funds from the European Union. IFAE is partially funded by the CERCA program of the Generalitat de Catalunya.
 


\bibliographystyle{mnras}
\bibliography{paper_draft} 

\section*{Affiliations}
{\small
$^{1}$Institute of Astronomy, University of Cambridge, Madingley Road, Cambridge, CB30HA, UK\\
$^{2}$Kavli Institute for Cosmology, University of Cambridge, Madingley Road, Cambridge, CB30HA, UK\\
$^{3}$Department of Astronomy, University of Illinois at Urbana-Champaign, Urbana, IL 61801, USA\\
$^{4}$National Center for Supercomputing Applications, University of Illinois at Urbana-Champaign, Urbana, IL 61801, USA\\
$^{5}$Department of Physics \& Astronomy, University College London, Gower Street, London, WC1E 6BT, UK\\
$^{6}$Department of Physics and Electronics, Rhodes University, PO Box 94, Grahamstown, 6140, South Africa\\
$^{7}$CNRS, UMR 7095, Institut d'Astrophysique de Paris, F-75014, Paris, France\\
$^{8}$Sorbonne Universit\'es, UPMC Univ Paris 06, UMR 7095, Institut d'Astrophysique de Paris, F-75014, Paris, France\\
$^{9}$Fermi National Accelerator Laboratory, P. O. Box 500, Batavia, IL 60510, USA\\
$^{10}$Institute of Cosmology \& Gravitation, University of Portsmouth, Portsmouth, PO1 3FX, UK\\
$^{11}$Laborat\'orio Interinstitucional de e-Astronomia - LIneA, Rua Gal. Jos\'e Cristino 77, Rio de Janeiro, RJ - 20921-400, Brazil\\
$^{12}$Observat\'orio Nacional, Rua Gal. Jos\'e Cristino 77, Rio de Janeiro, RJ - 20921-400, Brazil\\
$^{13}$Department of Astronomy, University of Illinois, 1002 W. Green Street, Urbana, IL 61801, USA\\
$^{14}$National Center for Supercomputing Applications, 1205 West Clark St., Urbana, IL 61801, USA\\
$^{15}$Institut de F\'{\i}sica d'Altes Energies (IFAE), The Barcelona Institute of Science and Technology, Campus UAB, 08193 Bellaterra (Barcelona) Spain\\
$^{16}$Kavli Institute for Particle Astrophysics \& Cosmology, P. O. Box 2450, Stanford University, Stanford, CA 94305, USA\\
$^{17}$Department of Physics and Astronomy, University of Pennsylvania, Philadelphia, PA 19104, USA\\
$^{18}$George P. and Cynthia Woods Mitchell Institute for Fundamental Physics and Astronomy, and Department of Physics and Astronomy, Texas A\&M University, College Station, TX 77843,  USA\\
$^{19}$Department of Physics, IIT Hyderabad, Kandi, Telangana 502285, India\\
$^{20}$Institut de Ci\`encies de l'Espai, IEEC-CSIC, Campus UAB, Carrer de Can Magrans, s/n,  08193 Bellaterra, Barcelona, Spain\\
$^{21}$Kavli Institute for Cosmological Physics, University of Chicago, Chicago, IL 60637, USA\\
$^{22}$Instituto de Fisica Teorica UAM/CSIC, Universidad Autonoma de Madrid, 28049 Madrid, Spain\\
$^{23}$Department of Astronomy, University of Michigan, Ann Arbor, MI 48109, USA\\
$^{24}$Department of Physics, University of Michigan, Ann Arbor, MI 48109, USA\\
$^{25}$SLAC National Accelerator Laboratory, Menlo Park, CA 94025, USA\\
$^{26}$Center for Cosmology and Astro-Particle Physics, The Ohio State University, Columbus, OH 43210, USA\\
$^{27}$Department of Physics, The Ohio State University, Columbus, OH 43210, USA\\
$^{28}$Astronomy Department, University of Washington, Box 351580, Seattle, WA 98195, USA\\
$^{29}$Cerro Tololo Inter-American Observatory, National Optical Astronomy Observatory, Casilla 603, La Serena, Chile\\
$^{30}$Santa Cruz Institute for Particle Physics, Santa Cruz, CA 95064, USA\\
$^{31}$Australian Astronomical Observatory, North Ryde, NSW 2113, Australia\\
$^{32}$Argonne National Laboratory, 9700 South Cass Avenue, Lemont, IL 60439, USA\\
$^{33}$Departamento de F\'isica Matem\'atica, Instituto de F\'isica, Universidade de S\~ao Paulo, CP 66318, S\~ao Paulo, SP, 05314-970, Brazil\\
$^{34}$Department of Astronomy, The Ohio State University, Columbus, OH 43210, USA\\
$^{35}$Instituci\'o Catalana de Recerca i Estudis Avan\c{c}ats, E-08010 Barcelona, Spain\\
$^{36}$Jet Propulsion Laboratory, California Institute of Technology, 4800 Oak Grove Dr., Pasadena, CA 91109, USA\\
$^{37}$Department of Physics and Astronomy, Pevensey Building, University of Sussex, Brighton, BN1 9QH, UK\\
$^{38}$Centro de Investigaciones Energ\'eticas, Medioambientales y Tecnol\'ogicas (CIEMAT), Madrid, Spain\\
$^{39}$School of Physics and Astronomy, University of Southampton,  Southampton, SO17 1BJ, UK\\
$^{40}$Instituto de F\'isica Gleb Wataghin, Universidade Estadual de Campinas, 13083-859, Campinas, SP, Brazil\\
$^{41}$Computer Science and Mathematics Division, Oak Ridge National Laboratory, Oak Ridge, TN 37831\\
}


\appendix
\section{MCMC FITTING} 

\begin{figure*}
	\centering    
    \subfigure[ULASJ0016-0038]{\label{fig:corner1}\includegraphics[width=0.32\textwidth]{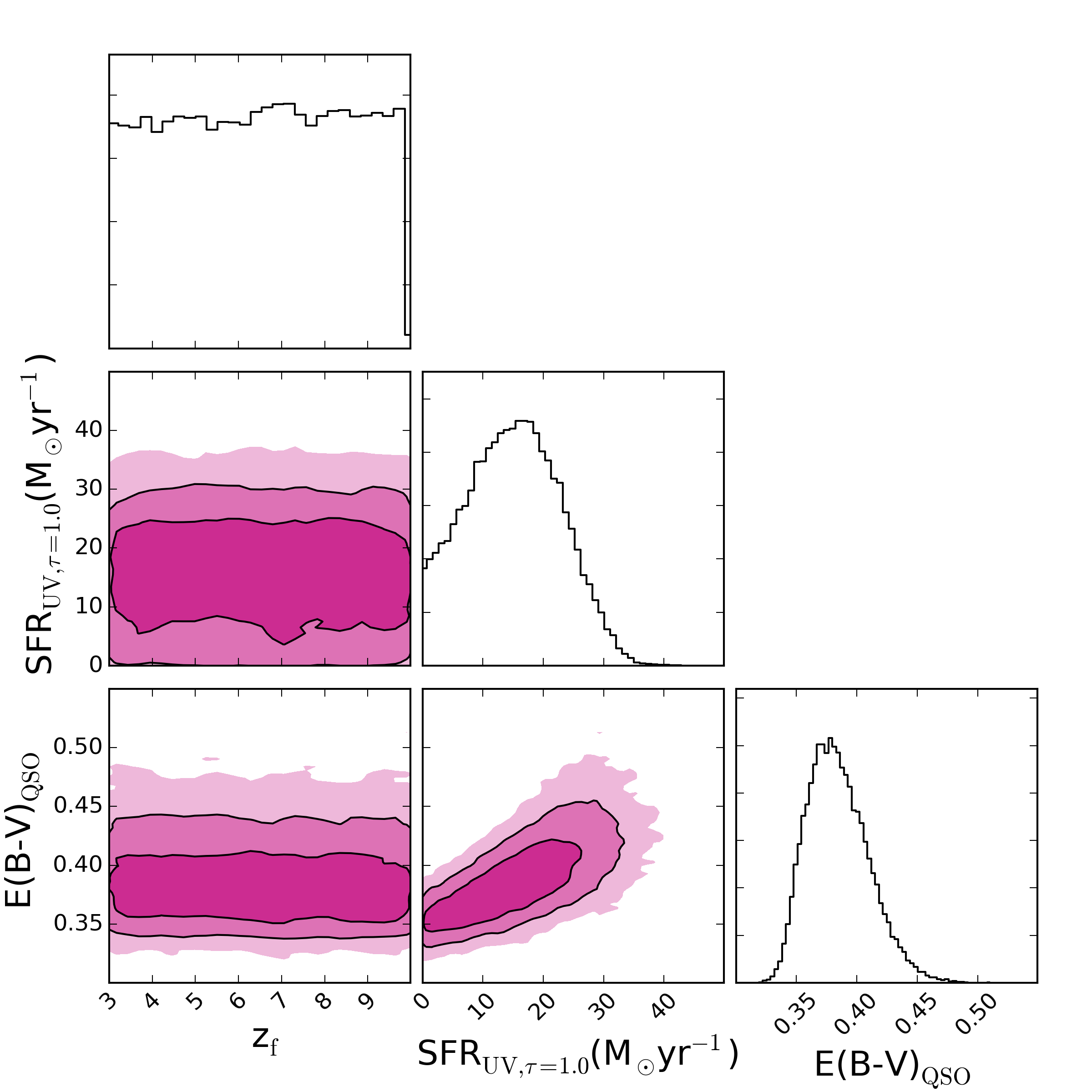}}
    \subfigure[ULASJ1002+0137]{\label{fig:corner2}\includegraphics[width=0.32\textwidth]{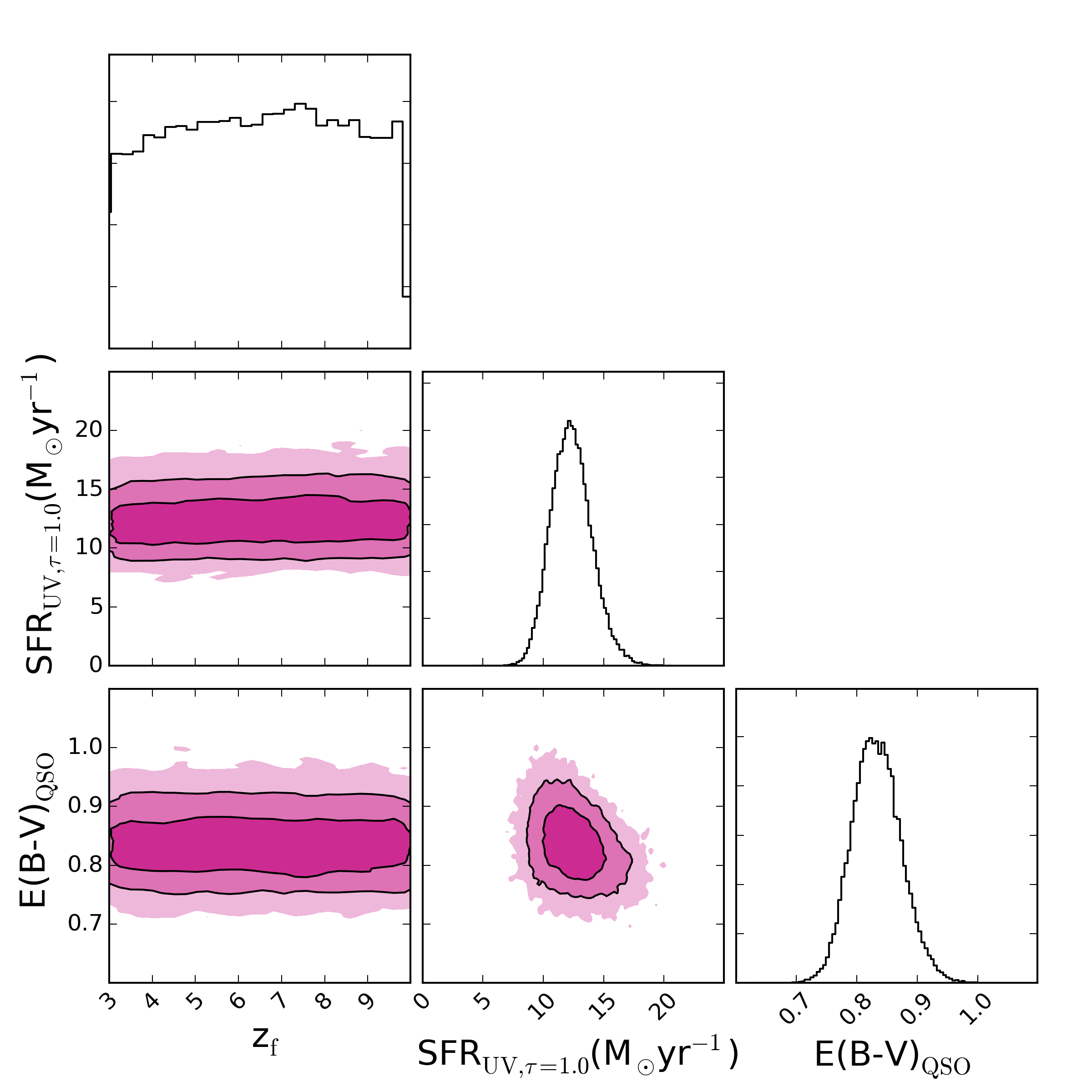}}
	\subfigure[VHSJ2024-5623]{\label{fig:corner3}\includegraphics[width=0.32\textwidth]{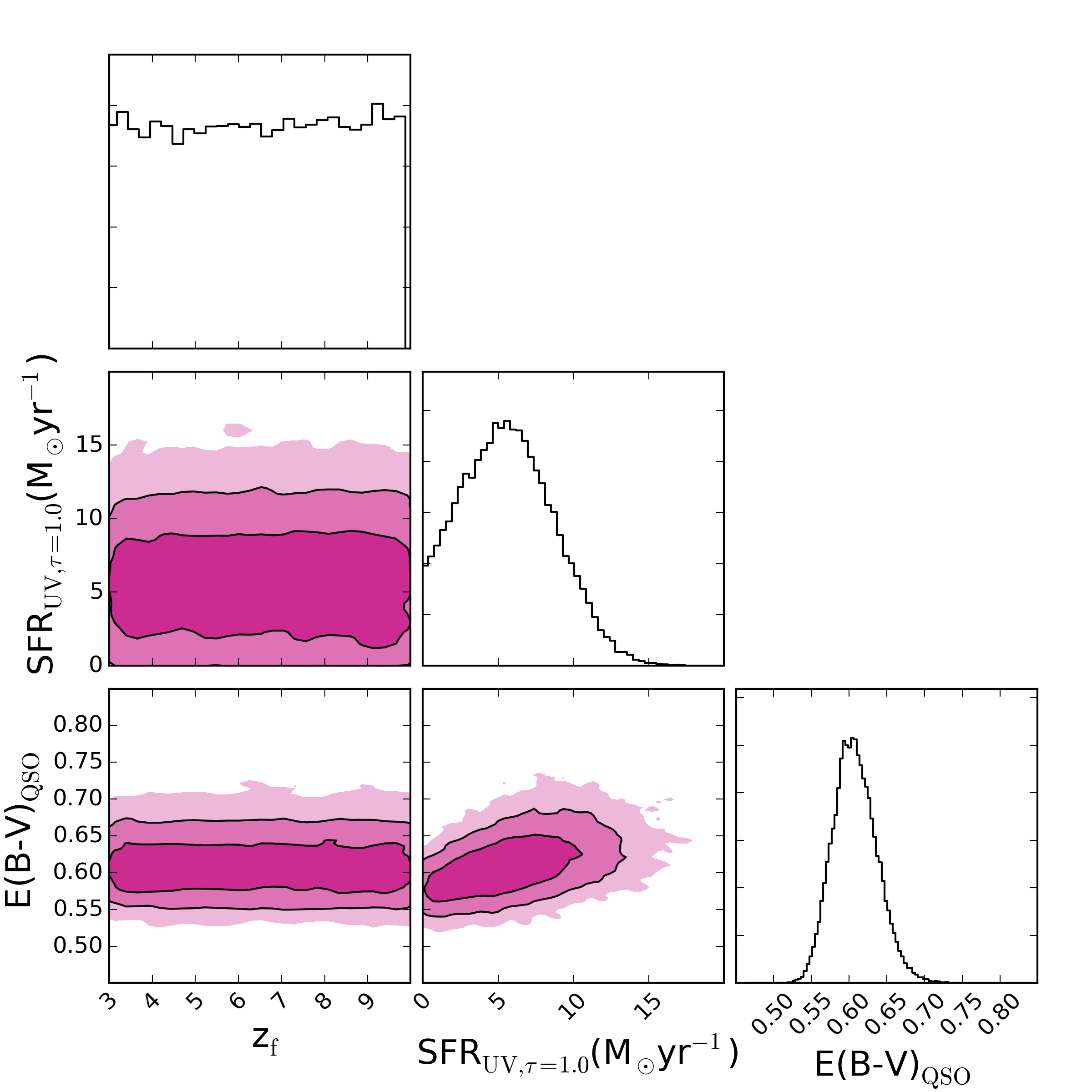}}
	\subfigure[VHSJ2028-5740]{\label{fig:corner4}\includegraphics[width=0.32\textwidth]{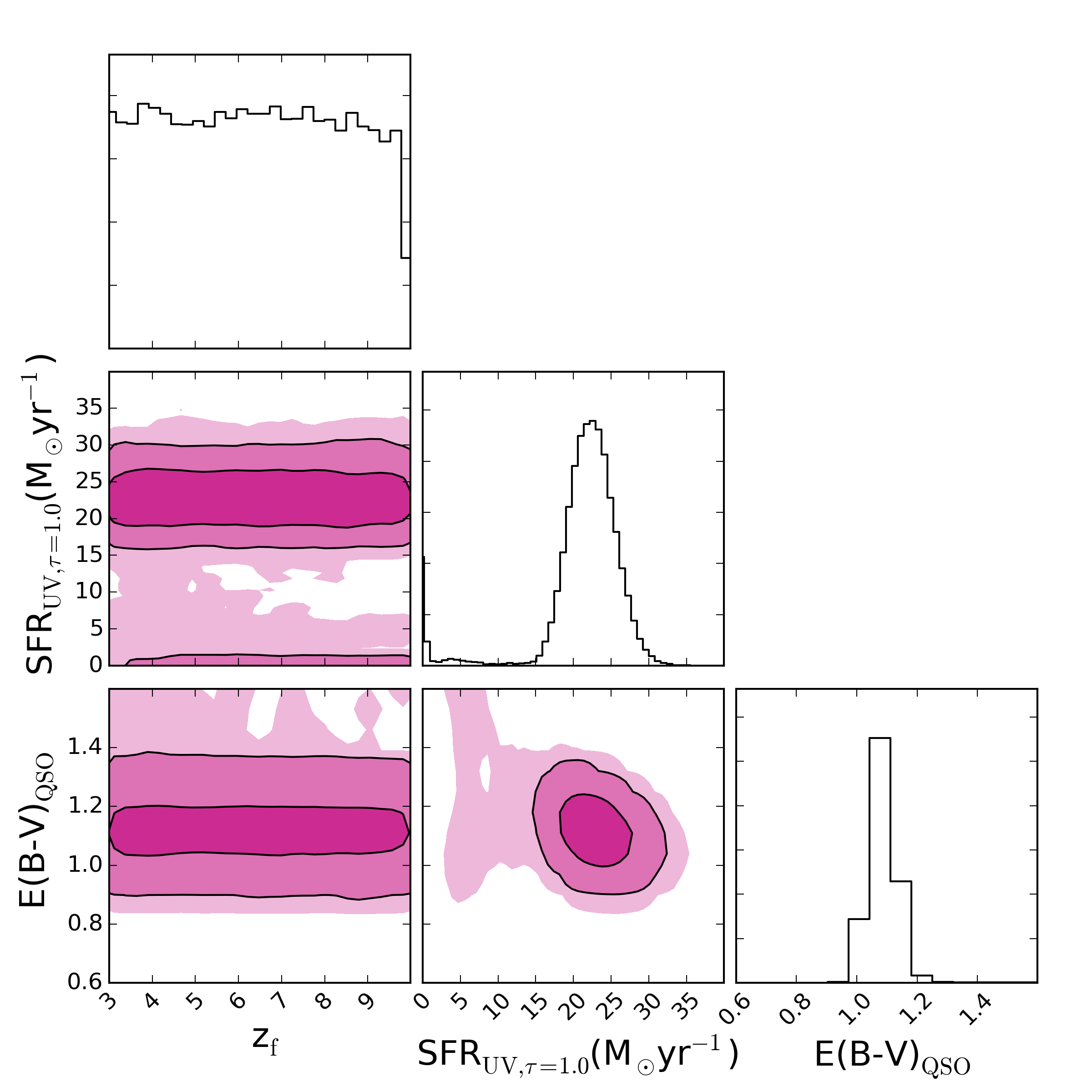}}
	\subfigure[VHSJ2100-5820]{\label{fig:corner5}\includegraphics[width=0.32\textwidth]{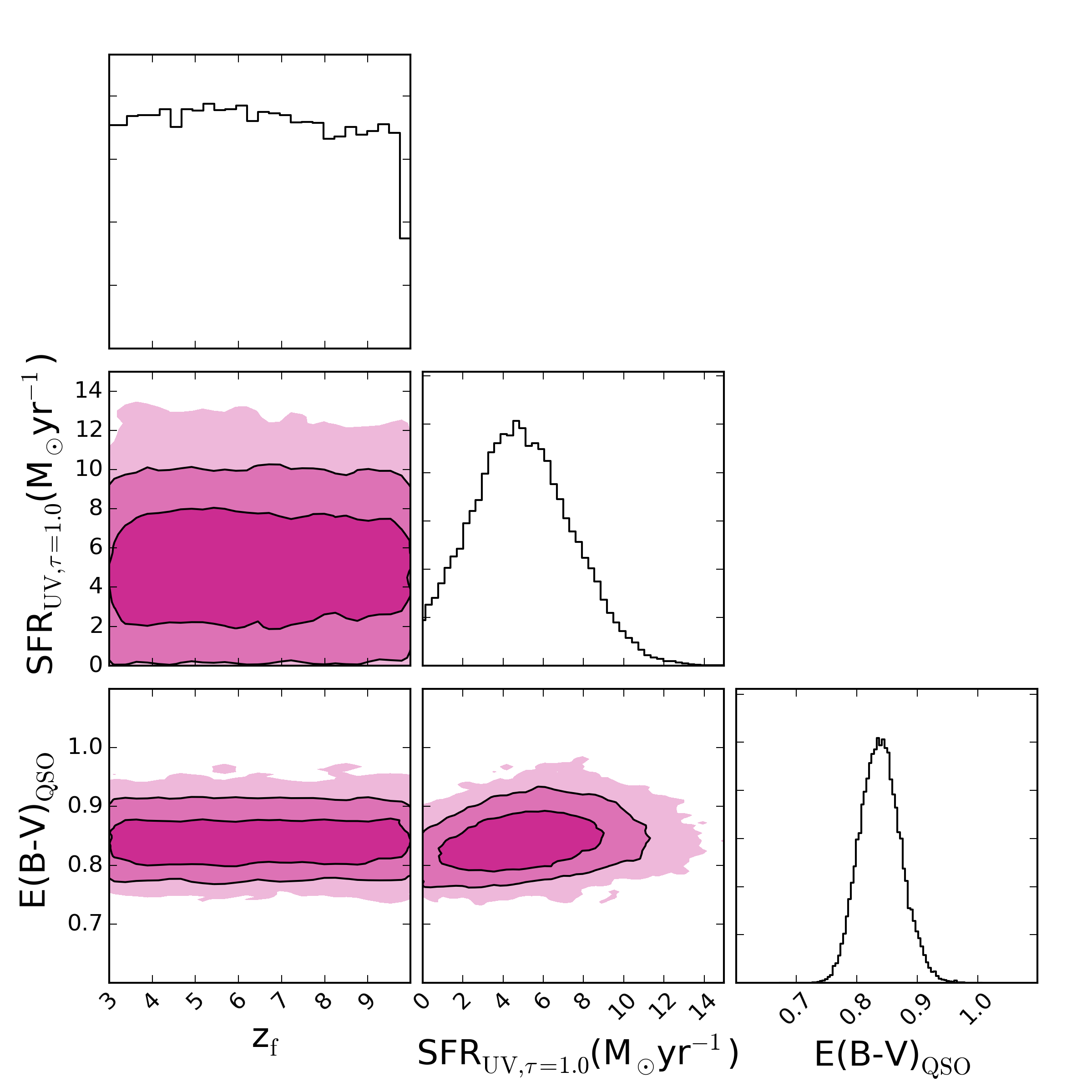}}
    \subfigure[VHSJ2115-5913]{\label{fig:corner6}\includegraphics[width=0.32\textwidth]{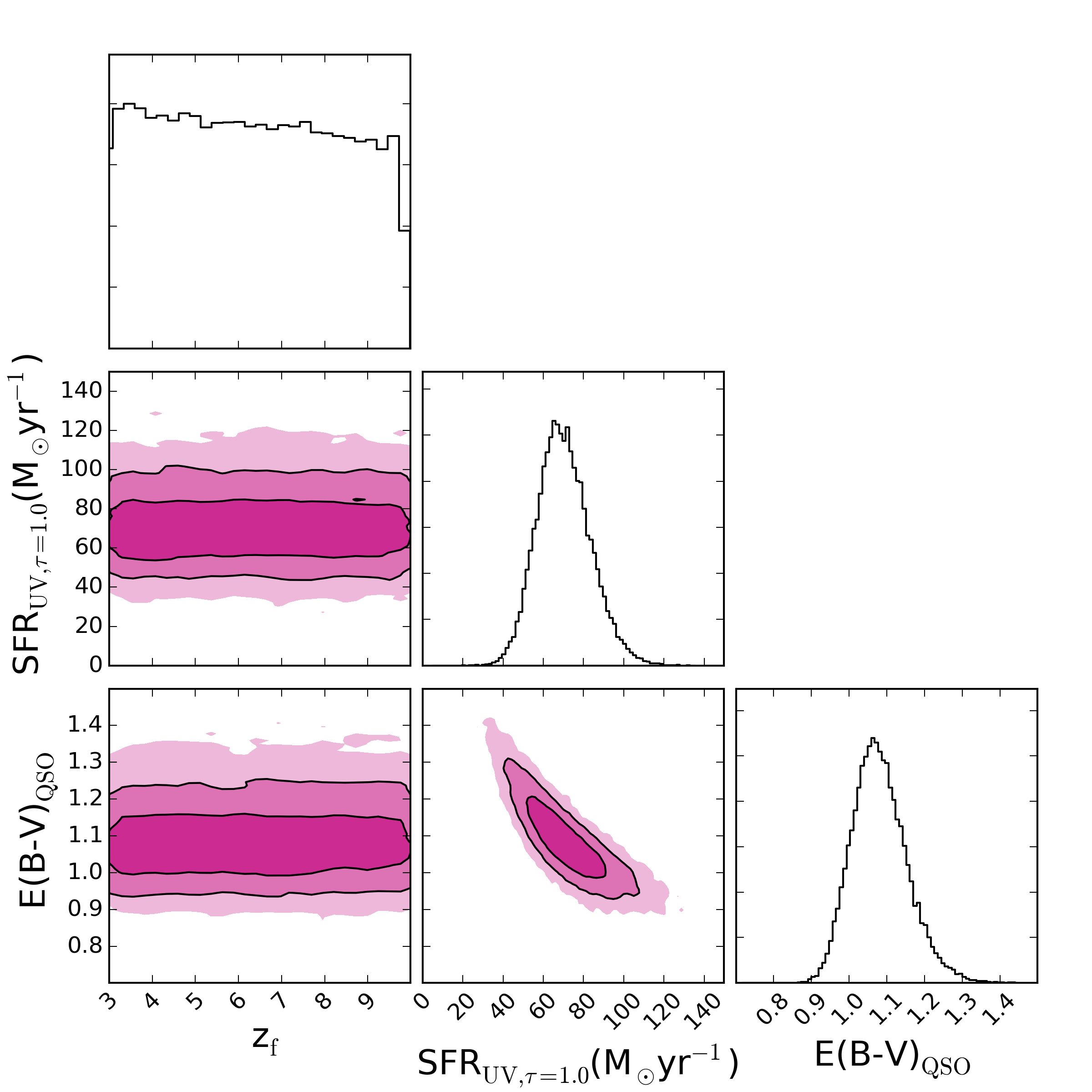}}
    \subfigure[ULASJ2200+0056]{\label{fig:corner7}\includegraphics[width=0.32\textwidth]{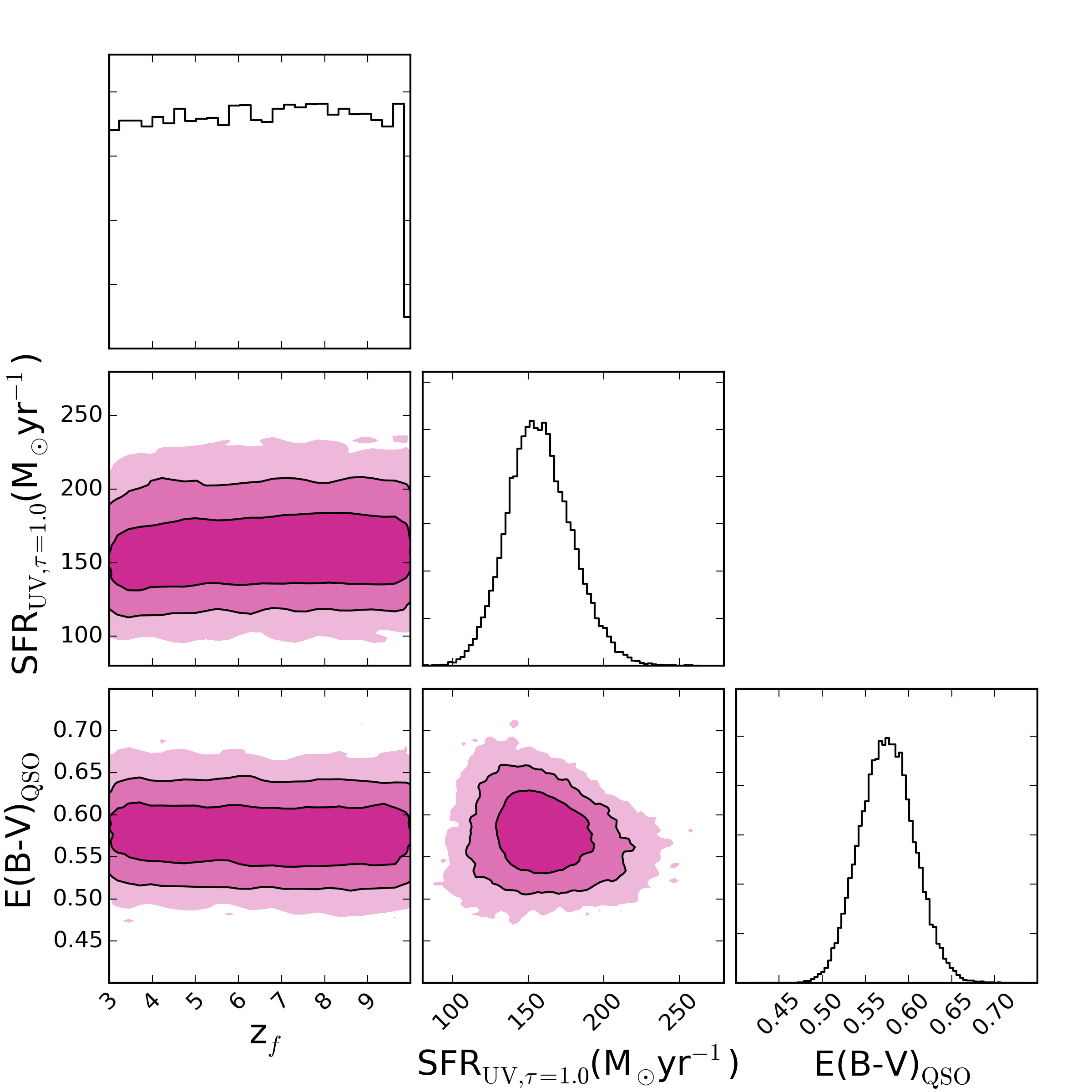}} 
    \subfigure[VHSJ2220-5618]{\label{fig:corner8}\includegraphics[width=0.32\textwidth]{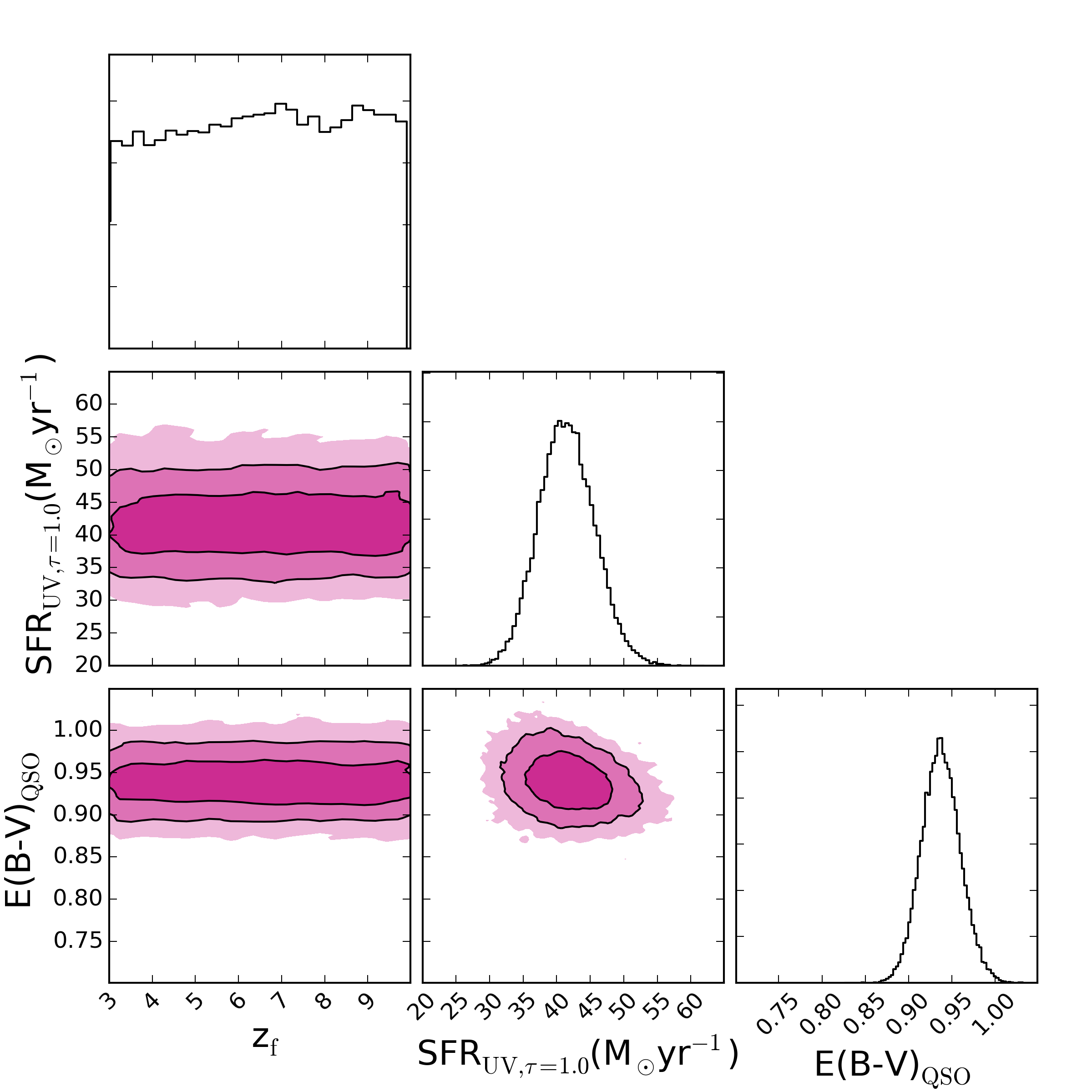}}
    \subfigure[ULASJ2224-0015]{\label{fig:corner9}\includegraphics[width=0.32\textwidth]{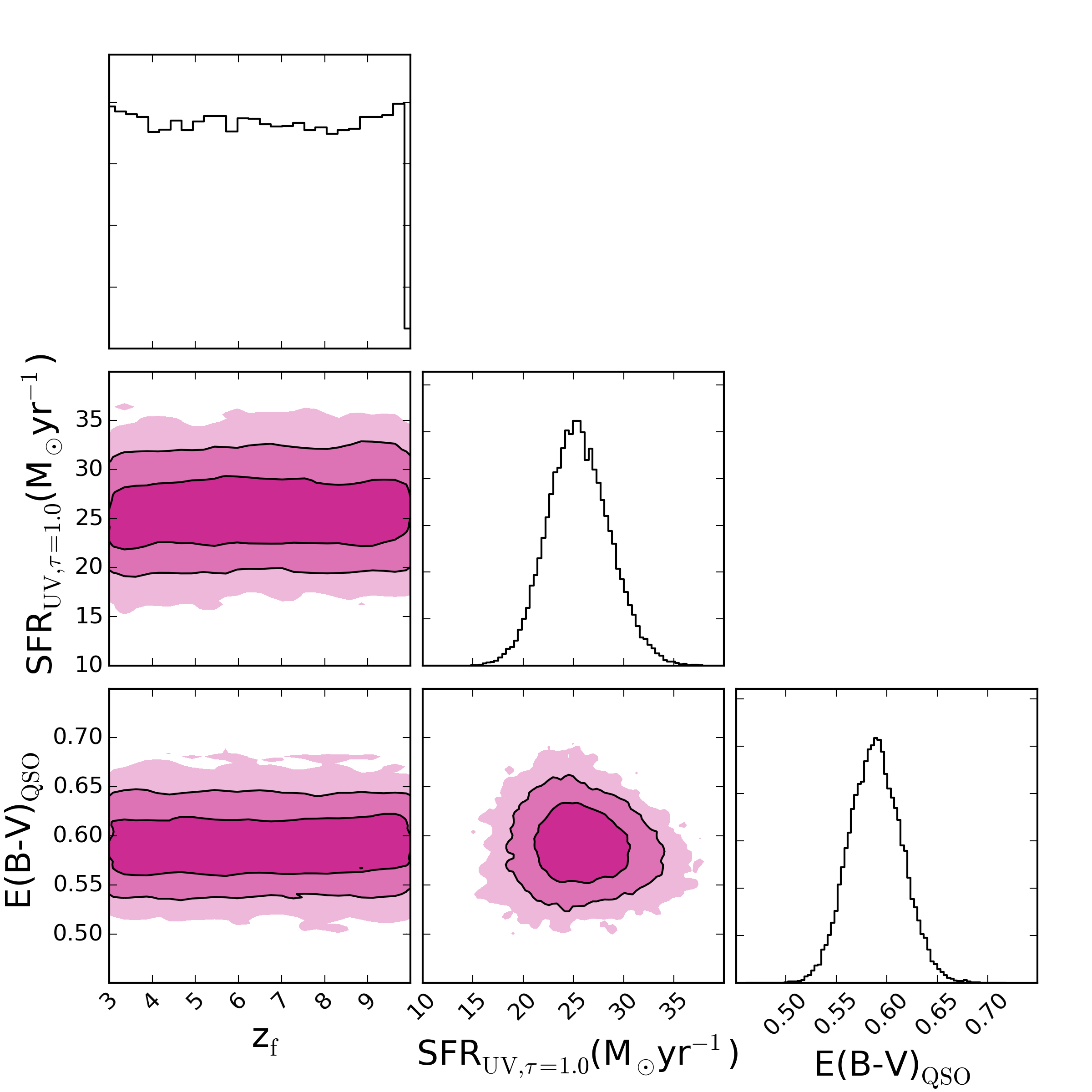}}
\caption{1D and 2D posterior distributions from the MCMC fitting. Shaded regions in the 2D distributions denote 1,2 and 3$\sigma$ parameter uncertainties in the fitting. SFR$_{\rm{UV,\tau_{v}=1.0}}$ and associated uncertainties (prior to dust corrections) are based on a galaxy template with $\tau_{\rm{V}}$ = 1.0 and have been converted from the normalisation of the galaxy template ($f_{\rm{gal}}$) in the fitting. Histograms illustrate the relative 1D probability distributions for each parameter.}
\label{fig:corner20}
\end{figure*}

\begin{figure*}
	\centering 
    \subfigure[VHSJ2227-5203]{\label{fig:corner10}\includegraphics[width=0.32\textwidth]{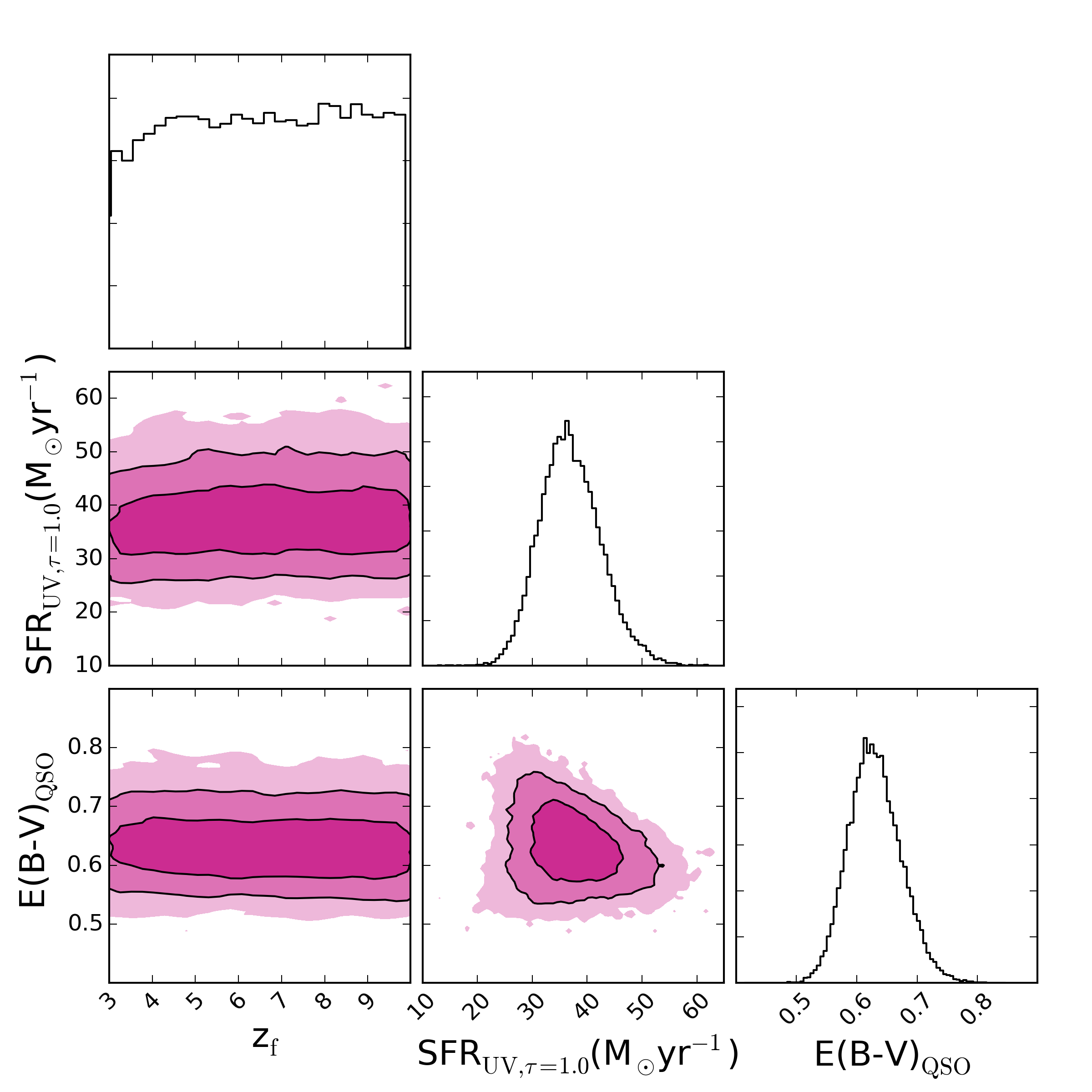}\addtocounter{subfigure}{9}}
    \subfigure[VHSJ2235-5750]{\label{fig:corner11}\includegraphics[width=0.32\textwidth]{VHSJ2235-5750.png}}
    \subfigure[VHSJ2256-4800]{\label{fig:corner12}\includegraphics[width=0.32\textwidth]{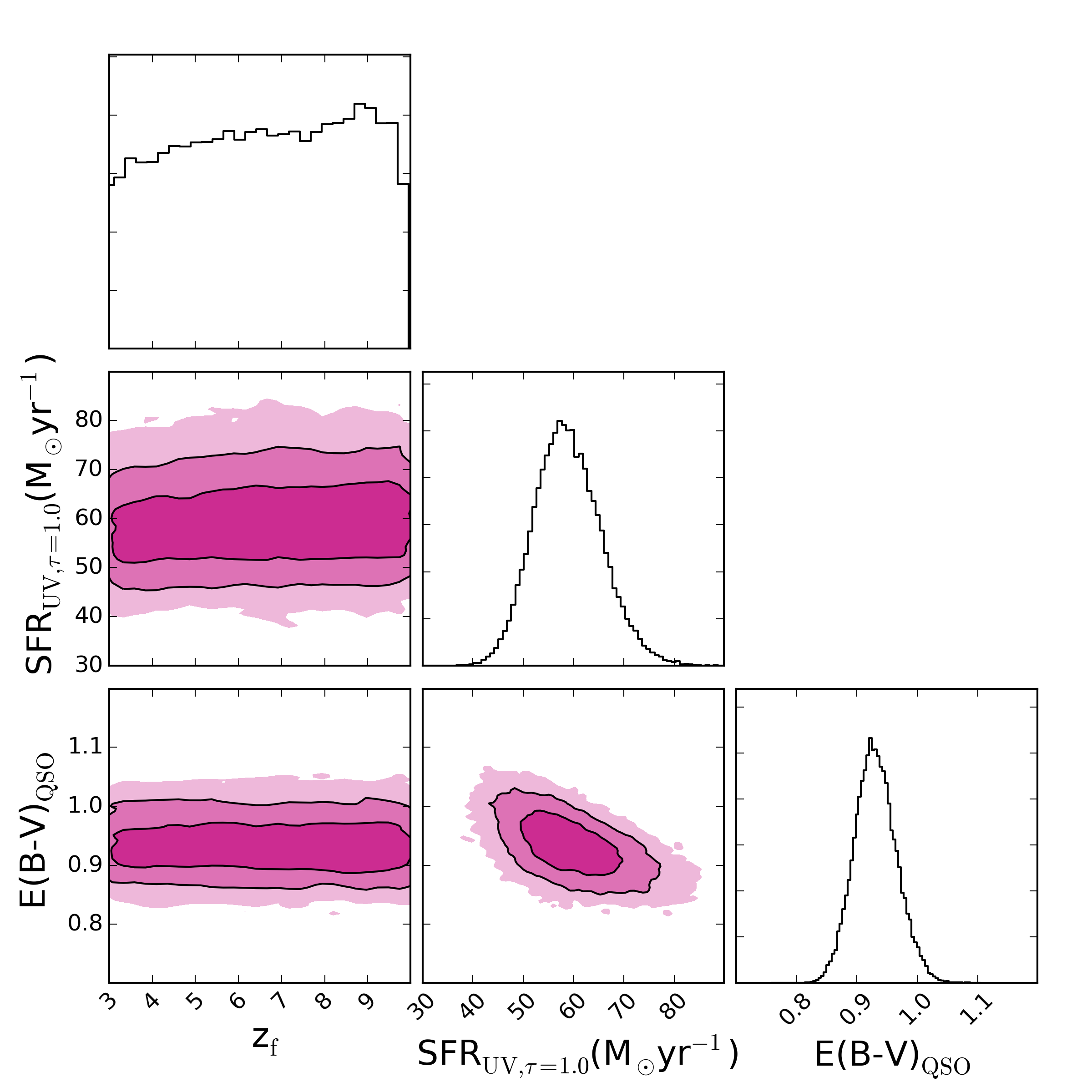}}
	\subfigure[VHSJ2257-4700]{\label{fig:corner13}\includegraphics[width=0.32\textwidth]{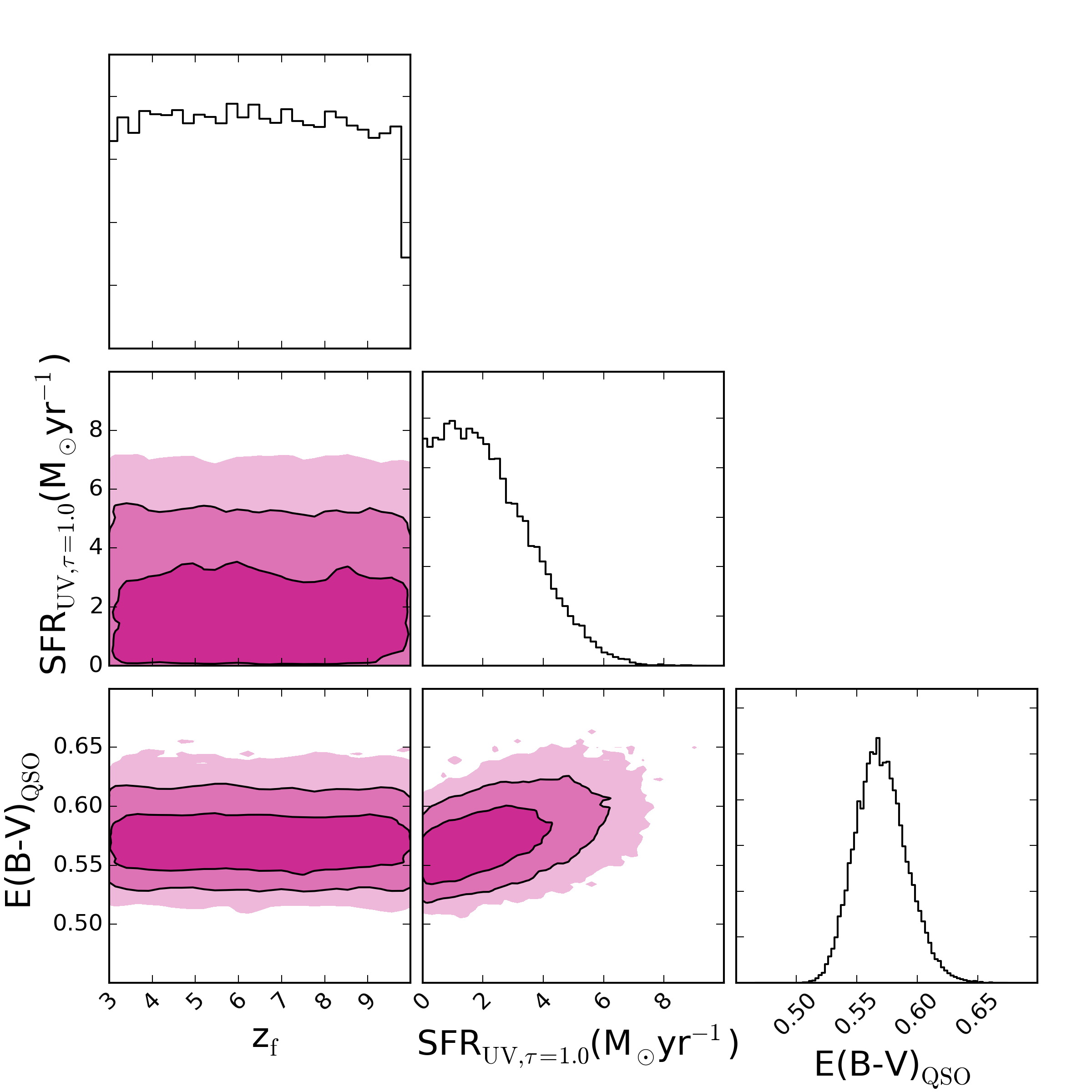}}
    \subfigure[VHSJ2306-5447]{\label{fig:corner14}\includegraphics[width=0.32\textwidth]{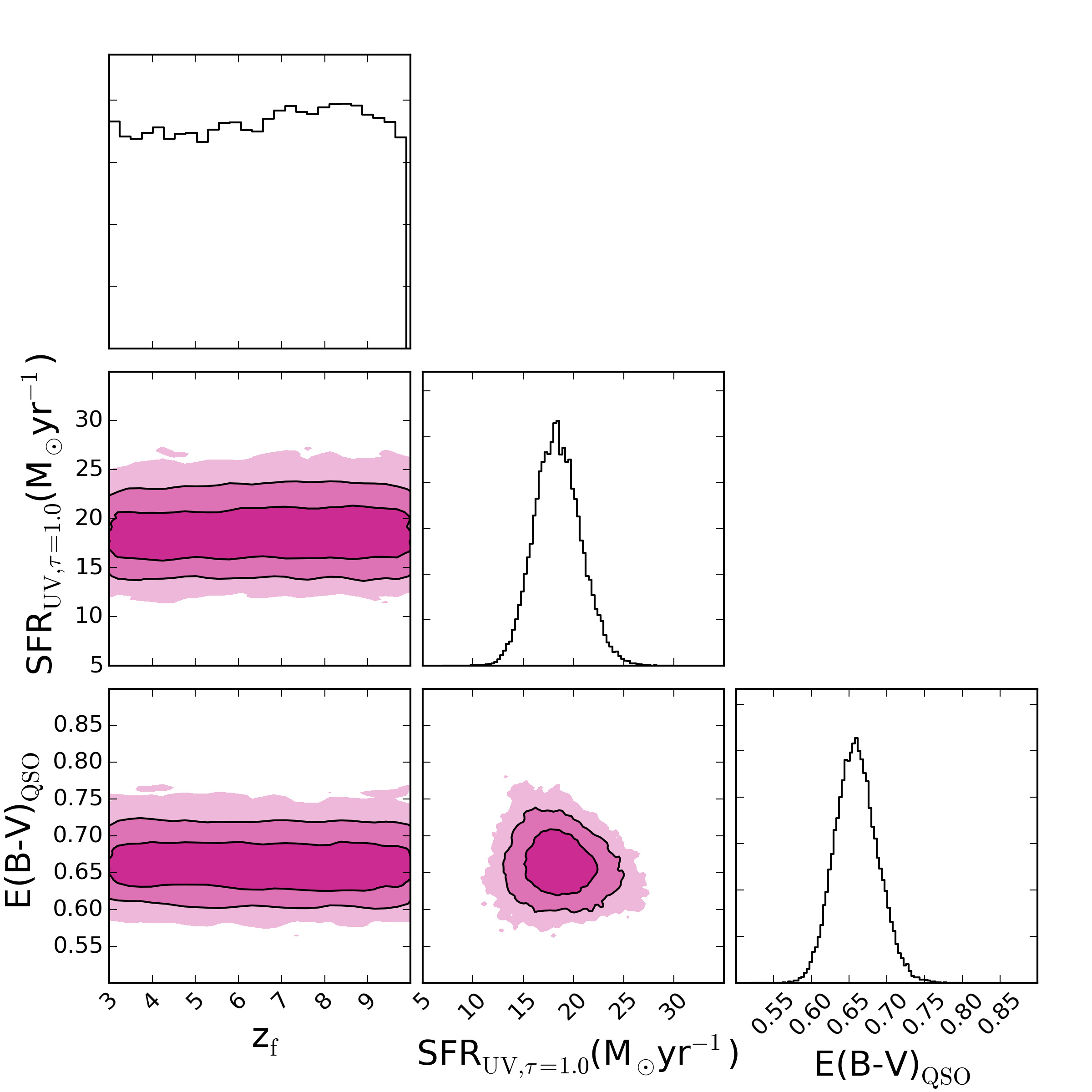}}
    \subfigure[ULASJ2315+0143]{\label{fig:corner15}\includegraphics[width=0.32\textwidth]{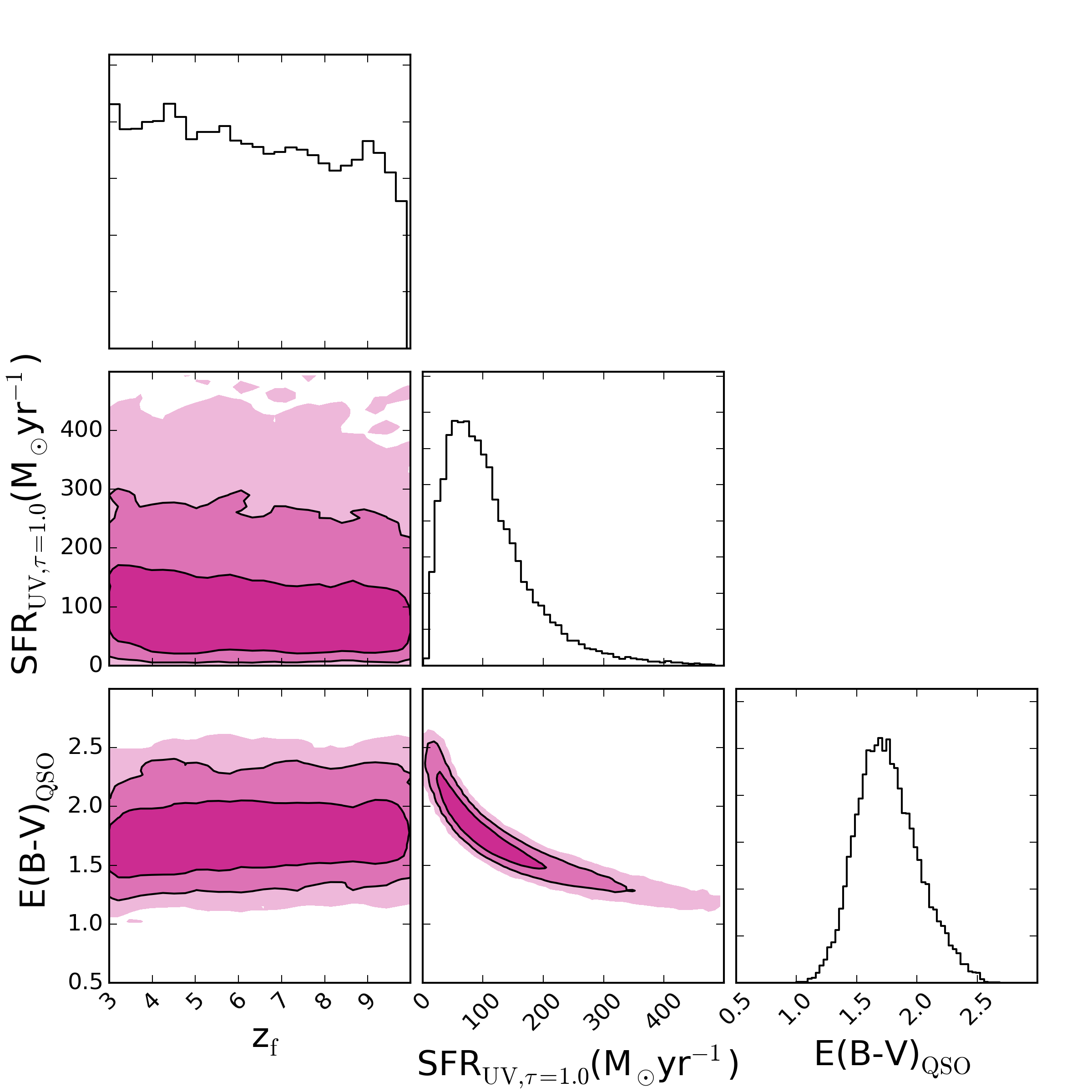}}
    \begin{center}
    \subfigure[VHSJ2332-5240]{\label{fig:corner16}\includegraphics[width=0.32\textwidth]{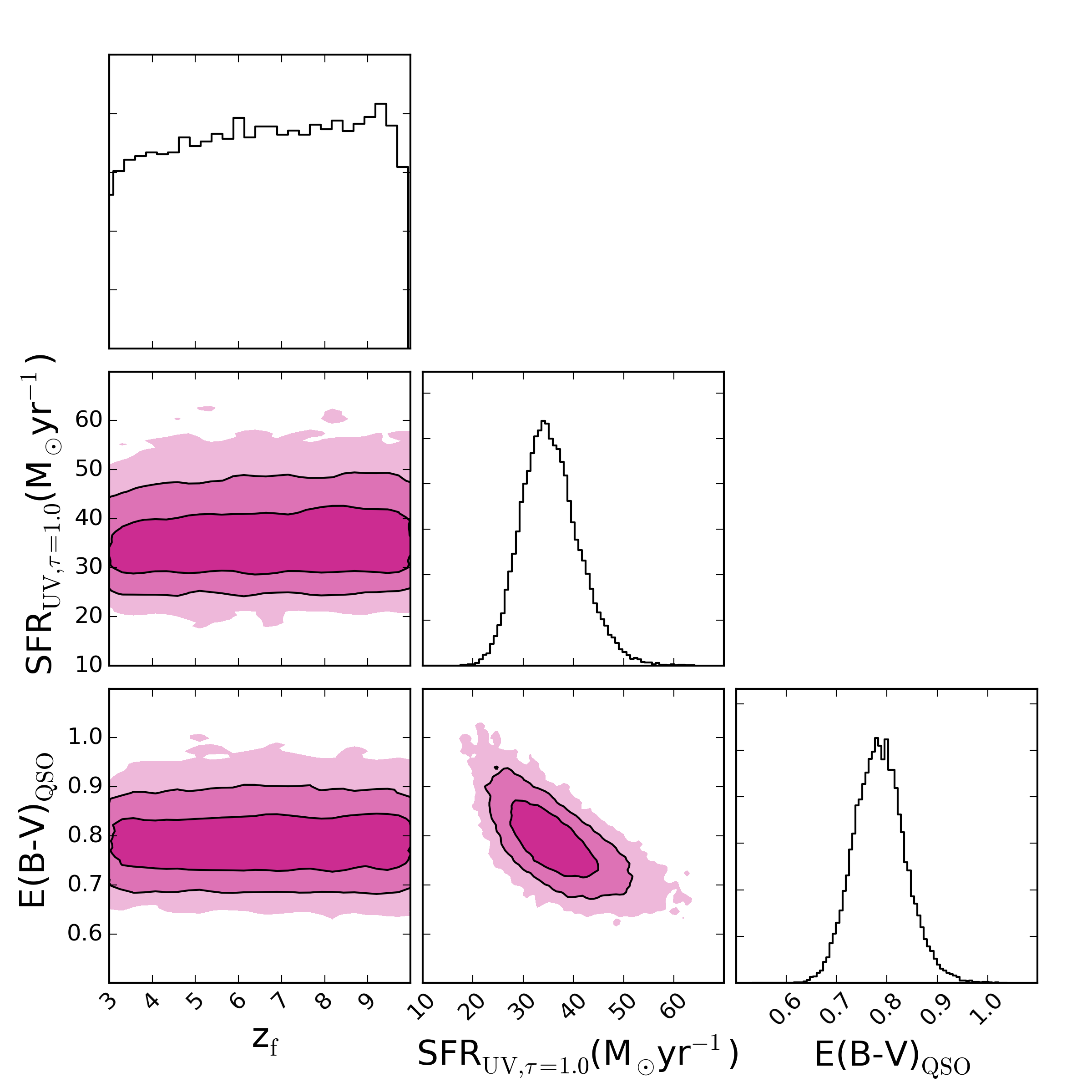}}
    \subfigure[VHSJ2355-0011]{\label{fig:corner17}\includegraphics[width=0.32\textwidth]{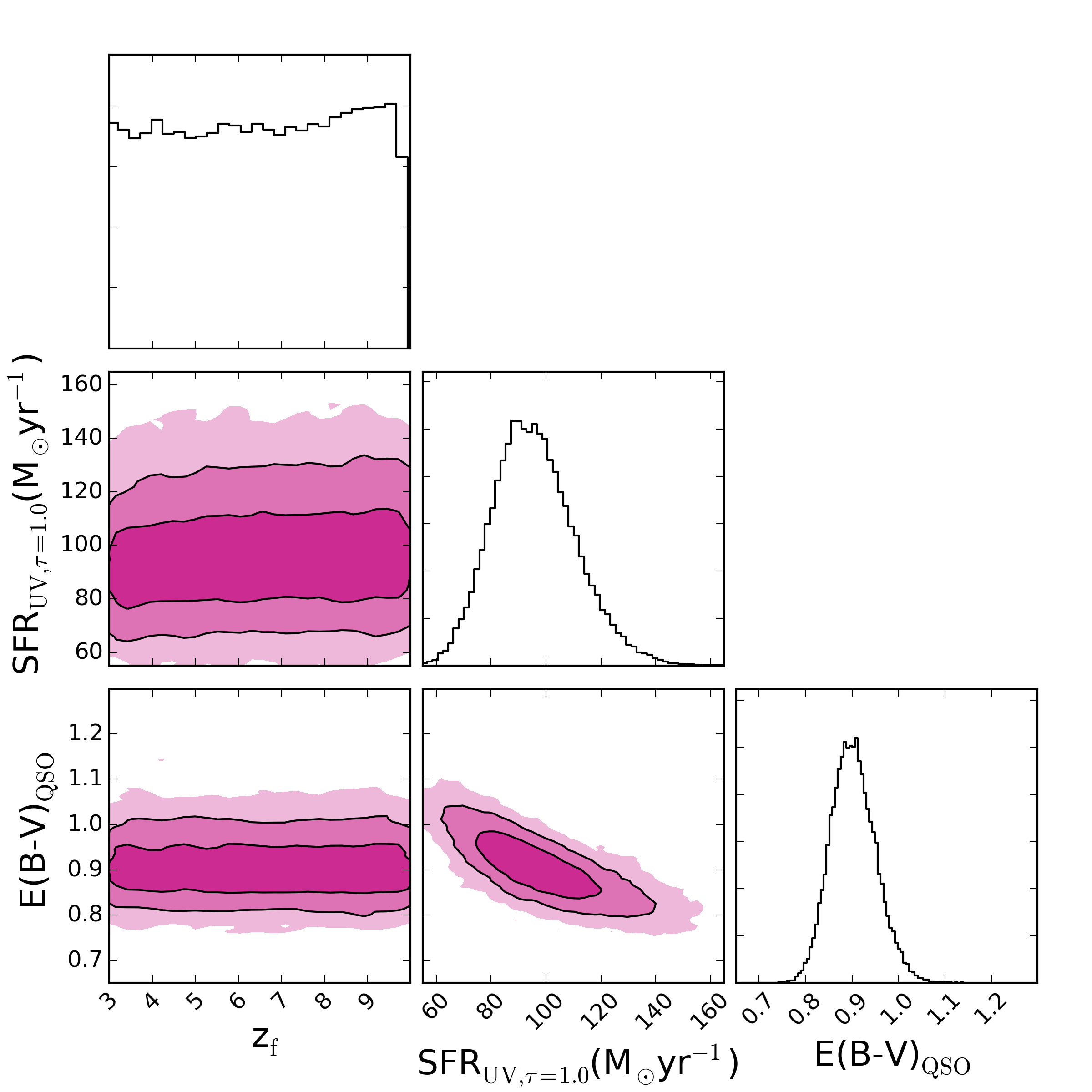}}
    \end{center}
\contcaption{}
\label{fig:corner21}
\end{figure*}

\bsp	
\label{lastpage}
\end{document}